\documentclass[american,english]{iopart}
\usepackage[T1]{fontenc}
\usepackage[latin9]{inputenc}
\usepackage{color}
\usepackage{babel}
\usepackage{array}
\usepackage{float}
\usepackage{multirow}
\usepackage{amsbsy}
\usepackage{amstext}
\usepackage{graphicx}
\usepackage[unicode=true,
 bookmarks=true,bookmarksnumbered=true,bookmarksopen=false,
 breaklinks=true,pdfborder={0 0 0},pdfborderstyle={},backref=false,colorlinks=true]
 {hyperref}
\hypersetup{
 pdfauthor={Edgar Garduno, Gabor T. Herman},
 pdfsubject={Image Reconstruction from Projections},
 pdfkeywords={Superiorization, Image Reconstruction, Shearlets, Regularization},
 linkcolor=blue,citecolor=blue,urlcolor=black}
\usepackage{breakurl}

\makeatletter

\providecommand{\tabularnewline}{\\}
\floatstyle{ruled}
\newfloat{algorithm}{tbp}{loa}
\providecommand{\algorithmname}{Algorithm}
\floatname{algorithm}{\protect\algorithmname}

\usepackage{iopams}
\usepackage{setstack}

\@ifundefined{date}{}{\date{}}
\usepackage[noadjust]{cite}
\usepackage{algorithm}
\usepackage{algorithmic}
\usepackage{graphicx}
\usepackage{color, colortbl}
\usepackage{relsize}
\usepackage[font=large,labelfont={normalsize}, textfont={normalsize,sf}]{subfig}

\@ifundefined{showcaptionsetup}{}{%
 \PassOptionsToPackage{caption=false}{subfig}}
\usepackage{subfig}
\makeatother

\addto\captionsamerican{\renewcommand{\algorithmname}{Algorithm}}

\begin{document}

\title{Computerized Tomography with Total Variation and with Shearlets}

\author{Edgar Gardu\~no\thanks{Edgar Gardu\~no is with Departamento de Ciencias de la Computación,
Instituto de Investigaciones en Matemáticas Aplicadas y en Sistemas,
Universidad Nacional Autónoma de México, Cd. Universitaria, C.P. 04510,
Mexico City, Mexico. E-mail: edgargar@ieee.org} and Gabor T. Herman\thanks{Gabor T. Herman is with Department of Computer Science, The Graduate
Center, City University of New York, New York, NY 10016, USA. E-mail:
gabortherman@yahoo.com}}
\begin{abstract}
To reduce the x-ray dose in computerized tomography (CT), many constrained
optimization approaches have been proposed aiming at minimizing a
regularizing function that measures lack of consistency with some
prior knowledge about the object that is being imaged, subject to
a (predetermined) level of consistency with the detected attenuation
of x-rays. One commonly investigated regularizing function is total
variation (TV), while other publications advocate the use of some
type of multiscale geometric transform in the definition of the regularizing
function, a particular recent choice for this is the shearlet transform.
Proponents of the shearlet transform in the regularizing function
claim that the reconstructions so obtained are better than those produced
using TV for texture preservation (but may be worse for noise reduction).
In this paper we report results related to this claim. In our reported
experiments using simulated CT data collection of the head, reconstructions
whose shearlet transform has a small $\ell_{1}$-norm are not more
efficacious than reconstructions that have a small TV value. Our experiments
for making such comparisons use the \textit{\emph{recently-developed
superiorization}} methodology for both regularizing functions. Superiorization
is an automated procedure for turning an iterative algorithm for producing
images that satisfy a primary criterion (such as consistency with
the observed measurements) into its superiorized version that will
produce results that, according to the primary criterion are as good
as those produced by the original algorithm, but in addition are superior
to them according to a secondary (regularizing) criterion. The method
presented for superiorization involving the $\ell_{1}$-norm of the
shearlet transform is novel and is quite general: It can be used for
any regularizing function that is defined as the $\ell_{1}$-norm
of a transform specified by the application of a matrix. Because in
the previous literature the split Bregman algorithm is used for similar
purposes, a section is included comparing the results of the superiorization
algorithm with the split Bregman algorithm.
\end{abstract}

\maketitle

\section{Introduction\label{sec:Introduction}}

In a typical x-ray computerized tomography (CT) study, projections
from various orientations are obtained by the scanner detectors and
later processed to produce a CT image that approximates, or \emph{reconstructs},
the internal distribution of the object's x-ray attenuation \cite{HERMAN:2009a}.
In recent years there has been an increased desire to reduce the x-ray
dosage in CT and, although there have been proposals to reduce the
x-ray radiation by decreasing either the current in the emitting x-ray
hardware or the duration of the x-ray pulse, the approach that has
received the most attention is the significant reduction of the number
of x-ray projections. Such schemes lead to a degradation of image
quality, in particular when the filtered back-projection (FBP) reconstruction
algorithm is used, the most common method to produce CT images \cite{PAN:2009a}.
Consequently, there has been a significant research effort on utilizing
constrained optimization approaches that aim at minimizing a regularizing
function that measures lack of consistency with some prior knowledge
about the nature of the object that is being imaged subject to (predetermined)
acceptable compatibility with the constraints provided by the detected
attenuation of x-rays.

In the literature there are many methods that employ various regularization
approaches; using total variation (TV) is a very popular choice, but
approaches using TV have been shown to have limitations when reconstructing
medically relevant images \cite{HERMAN:2008a}. Recently, it has been
suggested that a way to overcome the limitations of TV is by using
the $\ell_{1}$-norm of a sparse transform \cite{CHEN:2010a,YU:2010a}.
As an alternative to TV regularization, several authors have proposed
using wavelets, because they make possible to represent CT images
in a sparse manner \cite{ABASCAL:2012a}. More recently it has been
suggested that wavelets have limitations when representing objects
in two and three-dimensions; in particular, those objects that contain
edges \cite{CANDES:1999a,FENG:2009a,GAO:2011a,KUTYNIOK:2012a,YI:2009a}.
To deal with such drawbacks, some authors have proposed the use of
multiscale geometric analysis methods such as shearlets \cite{KUTYNIOK:2012a,VANDEGHINSTE:2013a}.
Shearlets form an affine system (i.e., they are obtained from a mother
shearlet by dilations, shears and translations). In the recently-published
work \cite{VANDEGHINSTE:2013a} it is reported that an algorithm that
uses for the regularizing function the $\ell_{1}$-norm of the shearlet
transform produces better results to preserve textural features than
an algorithm that uses TV as the regularizing function (but may be
not for reducing noise in the reconstructions).

In this paper we report on our investigation comparing the performance
of the $\ell_{1}$-norm of the shearlet transform with that of TV.
Based on simulated CT data of the human head, we report on cases in
which reconstructions whose shearlet transform has a small $\ell_{1}$-norm
are not more efficacious from the medical diagnosis point of view
than reconstructions that have a small TV value. We reach this conclusion
based on experiments that compare outputs produced by the recently-developed
superiorization methodology to the problem of CT reconstruction \cite{HERMAN:2012a,GARDUNO:2011a}
for both the $\ell_{1}$-norm of the shearlet transform and TV as
regularizing functions.

The superiorization methodology provides an automated process for
turning an iterative algorithm for producing images that are compatible
with the constraints provided by the measurements into its superiorized
version that will produce outputs that will be as good as those of
the original algorithm from the point of view of the primary criterion
of constraints compatibility, but will in addition also be good according
to a secondary criterion, such as the output having a low $\ell_{1}$-norm
of the shearlet transform or a low TV value. The superiorized algorithm
interlaces the iterative steps of the original algorithm for satisfying
the primary criterion with some steps, automatically determined by
the formula for the secondary criterion, that steer the process to
a solution appropriate for both of the criteria \cite{HERMAN:2008a,HERMAN:2012a,GARDUNO:2011a}.
To be more precise, before the application of the next step of the
original algorithm, the current iterate is perturbed so that it becomes
more desirable according to the secondary criterion. The automated
nature of this approach can save a lot of time and effort of a researcher
when faced with a new optimization task, since it does not require
the development of new mathematics for every new task.

The paper is organized as follows. The next section provides a brief
description of the CT reconstruction problem. Section \ref{sec:Superiorization}
provides a description of the superiorization methodology with both
regularization criteria. Section \ref{sec:Experiments} describes
the presented experiments and provides an analysis of their results.
The experiments include both an illustrative example and a study using
statistical hypothesis testing (SHT). They compare reconstructions
produced by applying the two above-mentioned relatively-new superiorization-based
algorithms and two more-classical algorithms applied to simulated
CT data. Section \ref{sec:split_Bregman_Method} presents an alternative
to the superiorization approach, namely the split Bregman method.
The final section contains a discussion and our conclusions.

\section{The CT Reconstruction Problem\label{sec:CT_Problem}}

For any real number $t$ and any angle $\theta\in\left[0,\pi\right)$,
we define the line integral $\left[\mathcal{R}f\right]\left(t,\theta\right)$
of the function $f$ of two real variables (representing the distribution
of the x-ray attenuation in a section of the object to be reconstructed)
in the direction $\theta$ at distance $t$ as 
\begin{equation}
\left[\mathcal{R}f\right]\left(t,\theta\right)=\int_{\mathbb{R}^{1}}f\left(t\cos\theta-s\sin\theta,t\sin\theta+s\cos\theta\right)ds.\label{eq:Line_Integral}
\end{equation}
This integral is commonly known as the \emph{ray} \emph{transform}.
Note that $\left(t\cos\theta,t\sin\theta\right)$ denotes the coordinates
of a point that is on the line along which we are integrating. A CT
scanner provides us with estimates of $\left[\mathcal{R}f\right]\left(t,\theta\right)$
for a finite collection of pairs $\left(t,\theta\right)$, this scanner-provided
information is frequently referred to as the \textit{projection data}.
We wish to recover the distribution of the x-ray attenuation from
the projection data; mathematically speaking, we wish to reconstruct
the function $f$ from a noisy and incomplete set of its line integrals
\cite{HERMAN:2009a,NATTERER:2001a}. (Note that this formulation is
specifically for the recovery of functions of two variables from their
estimated line integrals, but the presented approach is generalizable
to recovering functions of more than two variables.)

The methods for reconstructing functions from their projections (\emph{reconstruction
algorithms}) can be classified into two categories: transform-based
and series expansion methods. Transform-based methods take advantage
of the ray transform and its relationship to other transforms, such
as the Fourier transform, to provide closed-form solutions. Furthermore,
these methods treat the reconstruction problem as a continuous one
until the end, when an inversion formula is discretized. The series
expansion methods treat the reconstruction problem as a discrete problem
from the beginning. Transform-based methods are used when speed is
important and they are the most common method in commercial scanners
\cite{PAN:2009a}. On the other hand, series expansion methods have
gained renewed interest because of the desire to minimize radiation
dosage by reducing the size of projection data \cite{BEISTER:2012a}.
In this work we are interested in these types of algorithms because
they allow the specification of the sought-after reconstruction as
the solution of an optimization problem.

In CT it is typically assumed that the support of the function $f$
is subdivided into $J=M\times M$ small squares (called \emph{pixels}),
within which the value of the function is uniform; we use $x_{j}$
to denote this value within the $j$th of the $J$ pixels. Suppose
that measurements $y_{l}$ are made for $L$ lines, characterized
by $\left(t_{l},\theta_{l}\right)$, for $1\leq l\leq L$. This leads
to a system of approximate equations:
\begin{equation}
y_{l}\approx\sum_{j=1}^{J}r_{l,j}x_{j},\label{eq:y_l}
\end{equation}
where $r_{l,j}=\left[\mathcal{R}p_{j}\right]\left(t_{l},\theta_{l}\right)$
is the length of intersection of the $l$th line with the $j$th pixel.
There are published techniques in the literature for fast calculation
of the set of $r_{l,j}$; see, for example, \cite[Section 4.6]{HERMAN:2009a}.
An alternative notation for (\ref{eq:y_l}) is $\boldsymbol{y}\mathbf{\approx R}\boldsymbol{x}$.

This system of approximate equalities provides us with the constraints
that a proposed solution $\boldsymbol{x}$ ought to satisfy. For any
nonnegative real number $\varepsilon$, we say that a $J$-dimensional
vector $\boldsymbol{x}$ is $\varepsilon$-\textit{compatible} (with
the $L$-dimensional measurement vector $\boldsymbol{y}$ and the
$L\times J$ system matrix $\mathbf{R}$) if $\left\Vert \boldsymbol{y}-\mathbf{R}\boldsymbol{x}\right\Vert _{2}\leq\varepsilon$.
The $\ell_{2}$-norm $\left\Vert \boldsymbol{y}-\mathbf{R}\boldsymbol{x}\right\Vert _{2}$
is a \textit{proximity function} (see, for example, \cite{HERMAN:2012a})
that indicates by how much the proposed reconstruction $\boldsymbol{x}$
violates the constraints provided by the measurements taken by the
scanner. (More careful modeling of the underlying physical situation
leads to a proximity function with a weighted $\ell_{2}$-norm, which
is of the form $\left\Vert \mathbf{C}^{-1}\left(\boldsymbol{y}-\mathbf{R}\boldsymbol{x}\right)\right\Vert _{2}$,
where $\mathbf{C}$ is an $L\times L$ matrix used to model the detector
acquisition system and/or to compensate for errors due to noise in
the measurements \cite{HERMAN:2009a}; for example, the authors of
\cite{VANDEGHINSTE:2013a} \foreignlanguage{american}{chose a matrix}
$\mathbf{C}$ \foreignlanguage{american}{that reduced the weighting
of heavily attenuated rays with large relative uncertainty}. Our main
purpose in this paper is to compare the efficacy of using two different
regularization criteria in addition to $\varepsilon$-compatibilty.
Since we considered that for such a comparison the exact choice of
the matrix $\mathbf{C}$ may not be important, we decided to choose
$\mathbf{C}$ to be the identity matrix. This allows us to use the
unweighted norm and also to simplify the notation in all that follows
due to $\mathbf{C}$ being the identity. However, it is certainly
possible that this biases the conclusions based on our experimental
results. In particular, when comparing the efficacy of using for the
regularizing function the $\ell_{1}$-norm of the shearlet transform
as opposed to using TV, it may be the case that using the identity
for $\mathbf{C}$, rather than a matrix that models the detector acquisition
system more accurately, has a more negative effect for the shearlet
transform than for TV.)

From the practical point of view, an $\varepsilon$-compatible solution
is not necessarily a good one (even for a small $\varepsilon$), since
it does not take into consideration any prior knowledge about the
nature of the object that is being imaged. One approach to overcoming
this problem is by using a regularizing function $\phi$, such that
$\phi\left(\boldsymbol{x}\right)$ is an indicator of the prior undesirability
of a proposed reconstruction $\boldsymbol{x}$. With these considerations
in mind, the CT reconstruction problem can be reformulated as a constrained
optimization problem of the following kind: 
\begin{equation}
\mathbf{Find}\;\boldsymbol{x}^{*}=\arg\underset{\boldsymbol{x}}{\min}\,\phi\left(\boldsymbol{x}\right),\mathbf{\;subject\;to\;}\left\Vert \boldsymbol{y}-\mathbf{R}\boldsymbol{x}\right\Vert _{2}\leq\varepsilon.\label{eq:General_Optimization}
\end{equation}
There are many possible choices for the regularizing function $\phi$
of (\ref{eq:General_Optimization}).

A popular option is total variation (see, e.g., \cite{HERMAN:2008a,CANDES:2006a,CENSOR:2010a}),
which we define as follows. We index the pixels by $j$ and we let
$C$ denote the set of all indices of pixels that are not in the rightmost
column or the bottom row of the pixel array. For any pixel with index
$j$ in $C$, let $r\left(j\right)$ and $b\left(j\right)$ be the
index of the pixel to its right and below it, respectively. We define
TV by
\begin{equation}
\Upsilon\left(\boldsymbol{x}\right)=\sum_{j\in C}\sqrt{\left(x_{j}-x_{r\left(j\right)}\right)^{2}+\left(x_{j}-x_{b\left(j\right)}\right)^{2}}.\label{eq:total_variation}
\end{equation}

The authors of \cite{VANDEGHINSTE:2013a} proposed using the discrete
shearlet transform $\Psi$ and defining $\phi\left(\boldsymbol{x}\right)$
as $\left\Vert \Psi\left(\boldsymbol{x}\right)\right\Vert _{1}$;
i.e., the $\ell_{1}$-norm of $\Psi\left(\boldsymbol{x}\right)$.
This makes use of a directional multiscale framework that provides
a decomposition of a function over dilated, translated and orientated
versions of a fixed mother function; for details of the shearlets
transform and its implementation, we direct the readers to \cite{KUTYNIOK:2012a,LABATE:2005b,EASLEY:2006a,GOOSENS:2009a,GOOSENS:2011b,HAUSER:2014a}.
For our discussion here, the relevant observation is that there exists
an $I\times J$ matrix $\mathbf{S}$, such that $\mathbf{S}\boldsymbol{x}$
is the discrete shearlet transform of $\boldsymbol{x}$. For implementing
the discrete shearlet transform we use, following \cite{GOOSENS:2009a,GOOSENS:2011b},
the so-called \emph{Fast Non-Iterative Shearlet Transform} (with four
scales, eight orientations per scale, and 0.5 for the parameter controlling
both the bandwidth of the angular filters and the amount of redundancy
of the discrete shearlet transform, as suggested by the study in \cite{VANDEGHINSTE:2013a}).
We point out the potentially very useful fact that the method we present
for superiorizing for $\left\Vert \mathbf{S}\boldsymbol{x}\right\Vert _{1}$
does not depend on the actual components of the matrix $\mathbf{S}$
and so it is applicable to any other transform that can be defined
as a mapping of $\boldsymbol{x}$ into $\mathbf{S}\boldsymbol{x}$
for some matrix $\mathbf{S}$.

For both these choices of $\phi\left(\boldsymbol{x}\right)$ (namely,
$\Upsilon\left(\boldsymbol{x}\right)$ and $\left\Vert \mathbf{S}\boldsymbol{x}\right\Vert _{1}$),
we use the superiorization methodology to find an approximation to
the mathematically defined $\boldsymbol{x}^{*}$ that is the solution
of the constrained optimization problem (\ref{eq:General_Optimization}).

\section{The Superiorization Methodology\label{sec:Superiorization}}

As stated in Section \ref{sec:Introduction}, the superiorization
methodology \cite{HERMAN:2012a} is an automated process for turning
an iterative algorithm for producing images that are compatible with
the constraints provided by the measurements into its superiorized
version that will produce outputs that will be as good as those of
the original algorithm from the point of view of constraints compatibility,
but will in addition also be good according to a regularizing function.
Here we measure the constraints compatibility of a $J$-dimensional
vector $\boldsymbol{x}$, by $\left\Vert \boldsymbol{y}-\mathbf{R}\boldsymbol{x}\right\Vert _{2}$.
The algebraic reconstruction techniques (ART) form a particular class
of iterative algorithms for finding, given the $L$-dimensional measurement
vector $\boldsymbol{y}$ and the $L\times J$ system matrix $\mathbf{R}$,
a $J$-dimensional vector whose constraints compatibility is small
\cite{HERMAN:2009a,GORDON:1970a}.

A single iterative step of the particular version of ART that we use
in this paper is provided below by the procedure \textbf{\textsc{ART}}$\left(\mathbf{R},\boldsymbol{y},\boldsymbol{x},\boldsymbol{x'},\rho\right)$,
where $\mathbf{R}$ is an $L\times J$ (system) matrix, $\boldsymbol{y}$
is an $L$-dimensional (measurement) vector, $\boldsymbol{x}$ is
a $J$-dimensional (input) vector, $\boldsymbol{x'}$ is a $J$-dimensional
(output) vector and $\rho$ is a real number (called the \textit{relaxation
parameter}). For $1\leq l\leq L$, we use $\boldsymbol{r}_{l}$ to
denote the $J$-dimensional vector that is the transpose of the $l$th
row of $\mathbf{R}$ and $y_{l}$ to denote the $l$th component of
$\boldsymbol{y}$; recall (\ref{eq:y_l}). Following \cite[Ch. 11]{HERMAN:2009a},
the details of this procedure are:

\begin{algorithmic}[1]

\STATE{\textbf{procedure }\textbf{\textsc{ART}}$\left(\mathbf{R},\boldsymbol{y},\boldsymbol{x},\boldsymbol{x'},\rho\right)$}

\STATE{~~~~\textbf{set} $\boldsymbol{x'}\leftarrow\boldsymbol{x}$}

\STATE{~~~~\textbf{set} $l\leftarrow1$}

\STATE{~~~~\textbf{while} $l\leq L$}

\STATE{~~~~~~~~~\textbf{set} $\boldsymbol{x'}\leftarrow\boldsymbol{x'}+{\displaystyle \rho\frac{y_{l}-\left\langle \boldsymbol{r}_{l},\boldsymbol{x'}\right\rangle }{\left\langle \boldsymbol{r}_{l},\boldsymbol{r}_{l}\right\rangle }}\boldsymbol{r}_{l}$}

\STATE{~~~~~~~~~\textbf{set} $l\leftarrow l+1$}

\STATE{\textbf{end} \textbf{procedure}}

\end{algorithmic}To avoid numerical difficulties we assume that $\left\langle \boldsymbol{r}_{l},\boldsymbol{r}_{l}\right\rangle $
is bounded away from zero; in our implementation this is achieved
by removing from the system of approximate equations $\boldsymbol{y}\mathbf{\approx R}\boldsymbol{x}$
those for which $\left\langle \boldsymbol{r}_{l},\boldsymbol{r}_{l}\right\rangle <10^{-20}$.

As discussed in \cite[Ch. 11]{HERMAN:2009a}, if $0.0<\rho<2.0$,
repeated applications of procedure \textbf{\textsc{ART}}$\left(\mathbf{R},\boldsymbol{y},\boldsymbol{x},\boldsymbol{x'},\rho\right)$
can be used for finding an $\varepsilon$-compatible $J$-dimensional
vector, for a given large-enough $\varepsilon$. This can be achieved
by first setting $\boldsymbol{x}^{(0)}$ to an arbitrary $J$-dimensional
vector (in this paper we use the zero vector $\mathbf{0}$ as the
starting vector) and then repeatedly calling \textbf{\textsc{ART}}$\left(\mathbf{R},\boldsymbol{y},\boldsymbol{x}^{(k)},\boldsymbol{x}^{(k+1)},\rho\right)$
until we find an $\varepsilon$-compatible $\boldsymbol{x}^{(k)}$.
The number of iterations to get to such a vector depends on the ordering
of the rows of the matrix $\mathbf{R}$, in this paper we use the
so-called ``efficient ordering'' \cite[Section 11.4]{HERMAN:2009a}.
We refer to this entire process as the ``algorithm ART.''

We make use of the superiorization methodology, as published in \cite{HERMAN:2012a},
to turn the algorithm ART into its superiorized version, whose aim
is to produce an output that is also $\varepsilon$-compatible (just
as the output of unsuperiorized algorithm ART), but with the additional
property of having a second criterion much improved. Such a criterion
is specified by a function $\phi:\mathbb{R}^{J}\rightarrow\mathbb{R}$,
with the intention that an image in $\mathbb{R}^{J}$ for which the
value of $\phi$ is smaller is \textit{superior} (from the point of
view of the application at hand) to an image in $\mathbb{R}^{J}$
for which the value of $\phi$ is larger.

A general method for turning an iterative algorithm into such a superiorized
version is provided by the \textbf{Superiorized Version of Algorithm
P} in \cite{HERMAN:2012a}. The \textbf{Superiorized Version of ART
}that we provide below is just an adaptation of the \textbf{Superiorized
Version of Algorithm P} for the case when $\mathbf{P}$ is $\mathbf{ART}$
and for a regularizing function $\phi(\boldsymbol{x})$. The superiorized
version of ART depends on a specified initial image that we chose
to be the zero vector $\mathbf{0}$, the vector whose elements are
all zero, and a summable sequence $\left(\gamma_{\ell}\right)_{\ell=0}^{\infty}$
of positive real numbers (we choose $\gamma_{\ell}=\beta_{0}\alpha{}^{\ell}$,
where $\beta_{0}>0$ and $0<\alpha<1$). The algorithm also uses a
\{$true$, $false$\}-valued variable called $loop$; the inner $\mathbf{while}$
loop in the algorithm is executed while $loop$ is $true$. The user-specified
input parameters are the $\beta_{0}$, $\alpha$, $\rho$ (the relaxation
parameter used in ART), $N$ (an integer number), and $\varepsilon$
(the desired constraints compatibly).

\begin{algorithm}[H]
\caption{\label{alg:Superiorized_ART}\textbf{\textit{\emph{Superiorized Version
of ART}}}}
\end{algorithm}
\vspace{-5mm}

\begin{algorithmic}[1]

\STATE{\textbf{set} $k=0$}

\STATE{\textbf{set} $\boldsymbol{x}^{\left(0\right)}\leftarrow\mathbf{0}$}

\STATE{\textbf{set} $\ell\leftarrow-1$}

\STATE{\textbf{while }$\left\Vert \boldsymbol{y}-\mathbf{R}\boldsymbol{x}^{(k)}\right\Vert _{2}^{2}>\varepsilon$
}

\STATE{~~~~\textbf{set} $n\leftarrow0$}

\STATE{~~~~\textbf{set} $\boldsymbol{x}^{\left(k,n\right)}\leftarrow\boldsymbol{x}^{\left(k\right)}$}

\STATE{~~~~\textbf{while $n<N$}}

\STATE{\label{Line:NAV}~~~~~~~~\textbf{set $\boldsymbol{v}^{\left(k,n\right)}$
}to be a \textbf{\textit{nonascending}} vector for $\phi$ at $\boldsymbol{x}^{\left(k,n\right)}$}

\STATE{~~~~~~~~\textbf{set} \emph{$loop\leftarrow true$}}

\STATE{~~~~~~~~\textbf{while}\emph{ loop}}

\STATE{~~~~~~~~~~~~\textbf{set $\ell\leftarrow\ell+1$}}

\STATE{~~~~~~~~~~~~\textbf{set} $\beta\leftarrow\beta_{0}\times\alpha^{\ell}$}

\STATE{~~~~~~~~~~~~\textbf{set} $\boldsymbol{z}\leftarrow\boldsymbol{x}^{\left(k,n\right)}+\beta\boldsymbol{v}^{\left(k,n\right)}$}

\STATE{~~~~~~~~~~~~\textbf{if $\phi\left(\boldsymbol{z}\right)\leq\phi\left(\boldsymbol{x}^{\left(k\right)}\right)$
$\mathbf{then}$}}

\STATE{~~~~~~~~~~~~~~~~\textbf{set $n\leftarrow n+1$}}

\STATE{~~~~~~~~~~~~~~~~\textbf{set $\boldsymbol{x}^{\left(k,n\right)}\leftarrow\boldsymbol{z}$}}

\STATE{~~~~~~~~~~~~~~~~\textbf{set }\emph{$loop\leftarrow false$}}

\STATE{~~~~\textbf{call }\textbf{\textsc{ART}}$\left(\mathbf{R},\boldsymbol{y},\boldsymbol{x}^{\left(k,N\right)},\boldsymbol{x}^{\left(k+1\right)},\rho\right)$}

\STATE{~~~~\textbf{set $k\leftarrow k+1$}}

\STATE{\textbf{return} $\boldsymbol{x}^{\left(k\right)}$}

\end{algorithmic}

The essential idea of the \emph{superiorization methodology }presented
in \cite{HERMAN:2012a} is to \emph{perturb} the original iterative
process. In the Superiorized Version of ART above, the perturbation
is done in Steps 5-17, which produce the $\boldsymbol{x}^{\left(k,N\right)}$
that replaces (in Step 18) the $\boldsymbol{x}^{\left(k\right)}$
in the repeated calling of \textbf{\textsc{ART}}$\left(\mathbf{R},\boldsymbol{y},\boldsymbol{x}^{(k)},\boldsymbol{x}^{(k+1)},\rho\right)$
within the algorithm ART. These perturbations are considered \textit{bounded}
(see, e.g., Section II.C of \cite{HERMAN:2012a}) because it is the
case that
\begin{equation}
\boldsymbol{x}^{\left(k,N\right)}=\boldsymbol{x}^{\left(k\right)}+\beta_{k}\boldsymbol{v}^{\left(k\right)},\:\mathrm{for\:all\:}k\geq0,\label{eq:perturbations}
\end{equation}
where the sequence $\left(\beta_{k}\right)_{k=0}^{\infty}$ of nonnegative
real numbers is \textit{summable} (i.e., $\sum_{k=0}^{\infty}\beta_{k}\,<\infty$)
and the sequence $\left(\boldsymbol{v}^{\left(k\right)}\right)_{k=0}^{\infty}$
of vectors in $\mathbb{R}^{J}$ is bounded. Further, in order for
the algorithm to return an output $\boldsymbol{x}^{\left(k\right)}$
in Step 20 for which $\phi\left(\boldsymbol{x}^{\left(k\right)}\right)$
is small, the perturbations ought to be such that $\phi\left(\boldsymbol{x}^{\left(k,N\right)}\right)\leq\phi\left(\boldsymbol{x}^{\left(k\right)}\right)$,
for all $k\geq0$. In order to achieve satisfaction of this condition,
we make use of the concept of a vector $\boldsymbol{d}$ that is \emph{nonascending}
for $\phi$ at $\boldsymbol{x\in\mathbb{R}^{J}}$. According to the
definition in Section II.D in \cite{HERMAN:2012a}, such a vector
has the properties that $\left\Vert \boldsymbol{d}\right\Vert _{2}\leq1$
and there is a $\delta>0$ such that, for all $\lambda\in\left[0,\delta\right]$,
$\phi\left(\boldsymbol{x}+\lambda\boldsymbol{d}\right)\leq\phi\left(\boldsymbol{x}\right)$. 

The precise consequences of using bounded perturbations based on nonascending
vectors are discussed in \cite{HERMAN:2012a}. Roughly stated, the
results there imply that if $\varepsilon$ is large enough to ensure
that ART will find an $\varepsilon$-compatible vector, then the Superiorized
Version of ART will also return an $\varepsilon$-compatible vector,
but one for which the value of $\phi$ is likely to be much smaller
(and is never greater). These results depend on being able to find
(for Step 8 of the Superiorized Version of ART) a nonascending vector
for $\phi$ at $\boldsymbol{x}^{\left(k,n\right)}$. We make use of
the following consequence of Theorem 2 from \cite{HERMAN:2012a}.

\textsf{\textbf{Theorem}}\textsf{. Let $\phi:\mathbb{R}^{J}\rightarrow\mathbb{R}$
be a convex function and let $\boldsymbol{x}\in\mathbb{R}^{J}$. Let
$\boldsymbol{g}\in\mathbb{R}^{J}$ satisfy the property: For 1$\leq j\leq J$,
if the $j$th component $g_{j}$ of $\boldsymbol{g}$ is not zero,
then the partial derivative $\frac{\partial\phi}{\partial x_{j}}(\boldsymbol{x})$
of $\phi$ at $\boldsymbol{x}$ exists and its value is $g_{j}$.
Define $\boldsymbol{d}$ to be the zero vector if $\left\Vert \boldsymbol{g}\right\Vert =0$
and to be $-\boldsymbol{g}/\left\Vert \boldsymbol{g}\right\Vert $
otherwise. Then $\boldsymbol{d}$ is a nonascending vector for $\phi$
at $\boldsymbol{x}$.}

Below we compare the Superiorized Version of ART for two choices of
$\phi$, one based on TV ($\phi(\boldsymbol{x})=\Upsilon\left(\boldsymbol{x}\right)$,
see (\ref{eq:total_variation})) and the other based on shearlets
($\phi(\boldsymbol{x})=\left\Vert \mathbf{S}\boldsymbol{x}\right\Vert _{1}$).

\subsection{TV-Based Superiorization{\normalsize{}\label{sec:TV-Superiorization}}}

To generate the nonascending vector when $\phi(\boldsymbol{x})=\Upsilon(\boldsymbol{x})$
we make use of the details at the end of the Appendix of \cite{Garduno14},
based on Theorem 2 of \cite{HERMAN:2012a}, to specify below the procedure
\textbf{\textsc{NonascendingTV}}$\left(\boldsymbol{x},\boldsymbol{d},\zeta\right)$.
In this procedure $\boldsymbol{x}$ is a $J$-dimensional (input)
vector, $\boldsymbol{d}$ is a $J$-dimensional (output) vector (it
is a nonascending vector for $\phi$ at $\boldsymbol{x}$) and $\zeta$
is a user-specified (very small) positive real number whose purpose
is to avoid numerical difficulties caused by a division with a near-zero
number. We make use of a \{$true$, $false$\}-valued variable $unsafe$
that indicates a potential numerical difficulty. Recalling that we
use $r\left(j\right)$ and $b\left(j\right)$, respectively, to refer
to the indices of the pixels to the right and below the pixel with
index $j\in C$, we also introduce the notations $\ell\left(j\right)$
and $u\left(j\right)$ for the indices of the pixels to the left and
above (respectively) of the pixel with index $j$ and define $C_{1}$
(respectively, $C_{2}$) as the set of all indices of pixels that
are not in the leftmost column or the bottom row (respectively, the
top row or the rightmost column) of the pixel array. The procedure
computes the nonascending vector $\boldsymbol{g}$ of $\Upsilon$
at $\boldsymbol{x}$ by calculating its $j$th component $g_{j}$
as the partial derivative of $\Upsilon$ with respect to $x_{j}$.
It can be seen that there are at most three terms in the sum in (\ref{eq:total_variation})
involving $x_{j}$ and the partial derivative is the sum of the partial
derivatives with respect to $x_{j}$ of these terms, provided they
exist and do not cause numerical difficulties.

\begin{algorithmic}[1]

\STATE{\textbf{procedure }\textbf{\textsc{NonascendingTV}}$\left(\boldsymbol{x},\boldsymbol{d},\zeta\right)$}

\STATE{~~~~\textbf{set} $\boldsymbol{d}\leftarrow\mathbf{0}$}

\STATE{~~~~\textbf{set} $\boldsymbol{g}\leftarrow\mathbf{0}$}

\STATE{~~~~\textbf{for} $j=1$ to $J-1$ \textbf{do}}

\STATE{~~~~~~~~~\textbf{set} $unsafe\leftarrow false$}

\STATE{~~~~~~~~~\textbf{if} $j\in C$ \textbf{then}}

\STATE{~~~~~~~~~~~~\textbf{if} $\left(x_{j}-x_{r\left(j\right)}\right)^{2}+\left(x_{j}-x_{b\left(j\right)}\right)^{2}>\zeta$}

\STATE{~~~~~~~~~~~~~~~\textbf{then set $g_{j}\leftarrow g_{j}+\frac{2x_{j}-x_{r\left(j\right)}-x_{b\left(j\right)}}{\sqrt{\left(x_{j}-x_{r\left(j\right)}\right)^{2}+\left(x_{j}-x_{b\left(j\right)}\right)^{2}}}$}}

\STATE{~~~~~~~~~~~~~~~\textbf{else set} $unsafe\leftarrow true$}

\STATE{~~~~~~~~~\textbf{if} $j\in C_{1}$\textbf{ then}}

\STATE{~~~~~~~~~~~~\textbf{if} $\left(x_{\ell\left(j\right)}-x_{j}\right)^{2}+\left(x_{\ell\left(j\right)}-x_{b\left(\ell\left(j\right)\right)}\right)^{2}>\zeta$}

\STATE{~~~~~~~~~~~~~~~\textbf{then set $g_{j}\leftarrow g_{j}+\frac{x_{j}-x_{\ell\left(j\right)}}{\sqrt{\left(x_{\ell\left(j\right)}-x_{j}\right)^{2}+\left(x_{\ell\left(j\right)}-x_{b\left(\ell\left(j\right)\right)}\right)^{2}}}$}}

\STATE{~~~~~~~~~~~~~~~\textbf{else set} $unsafe\leftarrow true$}

\STATE{~~~~~~~~~\textbf{if} $j\in C_{2}$ \textbf{then}}

\STATE{~~~~~~~~~~~~\textbf{if} $\left(x_{u\left(j\right)}-x_{r\left(u\left(j\right)\right)}\right)^{2}+\left(x_{u\left(j\right)}-x_{j}\right)^{2}>\zeta$}

\STATE{~~~~~~~~~~~~~~~\textbf{then set $g_{j}\leftarrow g_{j}+\frac{x_{j}-x_{u\left(j\right)}}{\sqrt{\left(x_{u\left(j\right)}-x_{r\left(u\left(j\right)\right)}\right)^{2}+\left(x_{u\left(j\right)}-x_{j}\right)^{2}}}$}}

\STATE{~~~~~~~~~~~~~~~\textbf{else set} $unsafe\leftarrow true$}

\STATE{~~~~~~~~~\textbf{if} $unsafe$ \textbf{then set}
$g_{j}\leftarrow0$ }

\STATE{~~~~\textbf{if} $\left\Vert \boldsymbol{g}\right\Vert _{2}>\zeta$
\textbf{then}}

\STATE{~~~~~~~\textbf{set} $\boldsymbol{d}\leftarrow-\left(\boldsymbol{g}/\left\Vert \boldsymbol{g}\right\Vert _{2}\right)$}

\STATE{\textbf{end} \textbf{procedure}}

\end{algorithmic}

The TV-Based Superiorized Version of ART makes use of the above procedure
by calling \textbf{\textsc{NonascendingTV$\left(\boldsymbol{x}^{\left(k,n\right)},\boldsymbol{v}^{\left(k,n\right)},\zeta\right)$}}
in Step \ref{Line:NAV} of Algorithm \ref{alg:Superiorized_ART}.
It follows from Theorem 2 of \cite{HERMAN:2012a} that, for any positive
real number $\zeta$, the \textbf{$\boldsymbol{v}^{\left(k,n\right)}$}
that is returned by such a call will be a nonascending vector for
$\Upsilon$ (i.e., for TV) at $\boldsymbol{x}^{\left(k,n\right)}$.

\subsection{{\normalsize{}Shearlet-Based Superiorization\label{sec:Shearlet-Superiorization}}}

To obtain the nonascending vector when $\phi(\boldsymbol{x})=\left\Vert \mathbf{S}\boldsymbol{x}\right\Vert _{1}$,
we observe that 
\begin{equation}
\phi(\boldsymbol{x})={\displaystyle \sum_{i=1}^{I}\left|\sum_{j=1}^{J}s_{ij}x_{j}\right|}\label{eq:fi_in_PQ}
\end{equation}
is a convex function. In order to be able to apply the Theorem stated
above for obtaining (without numerical difficulties) a nonascending
vector, we need to avoid regions in which $\sum_{j=1}^{J}s_{ij}x_{j}$
is near zero for some $i$, $1\leq i\leq I$. We select a small positive
real number $\zeta$ and, for any $\boldsymbol{x}\in\mathbb{R}^{J}$,
we define the sets 
\begin{equation}
\begin{array}{ccc}
P_{\zeta}(\boldsymbol{x}) & = & \left\{ i\,|\,1\leq i\leq I\;\mathrm{and\;}{\displaystyle \sum_{j=1}^{J}}s_{ij}x_{j}>\zeta\right\} ,\\
N_{\zeta}(\boldsymbol{x}) & = & \left\{ i\,|\,1\leq i\leq I\;\mathrm{and\;}{\displaystyle \sum_{j=1}^{J}}s_{ij}x_{j}<-\zeta\right\} ,\\
Z_{\zeta}(\boldsymbol{x}) & = & \left\{ i\,|\,1\leq i\leq I\;\mathrm{and\;}\left|{\displaystyle \sum_{j=1}^{J}}s_{ij}x_{j}\right|\leq\zeta\right\} .
\end{array}\label{eq:PNZ_indices}
\end{equation}
Based on these sets, we see that the Theorem provides us with a nonascending
vector by using $\boldsymbol{g}=\left(g_{1},g_{2},\ldots,g_{J}\right)^{T}\in\mathbb{R}^{J}$
with
\begin{equation}
g_{j}=\left\{ \begin{array}{lc}
0, & \textnormal{if }Z_{\zeta}(\boldsymbol{x})\neq\emptyset,\\
\left({\displaystyle \sum_{i\epsilon P_{\zeta}(\boldsymbol{x})}}s_{ij}\right)-\left({\displaystyle \sum_{i\epsilon N_{\zeta}(\boldsymbol{x})}}s_{ij}\right), & \textnormal{otherwise}.
\end{array}\right.\label{eq:g_j}
\end{equation}

This leads us to the following procedure for obtaining a nonascending
vector for $\phi$ at a point $\boldsymbol{x}$:

\begin{algorithmic}[1]

\STATE{\textbf{procedure }\textbf{\textsc{NonascendingShearlet}}$\left(\boldsymbol{x},\boldsymbol{d},\zeta\right)$}

\STATE{~~~~\textbf{set} $cont\leftarrow true$}

\STATE{~~~~\textbf{set }$\boldsymbol{d}\leftarrow\boldsymbol{0}$}

\STATE{~~~~\textbf{set }$\boldsymbol{a}\leftarrow\mathbf{S}\boldsymbol{x}$}

\STATE{~~~~\textbf{set }$i\leftarrow0$}

\STATE{~~~~\textbf{while} $i<I$ and $cont=true$ \textbf{do}}

\STATE{~~~~~~~~~\textbf{set} $i\leftarrow i+1$}

\STATE{~~~~~~~~~\textbf{if} $\left|a_{i}\right|\leq\zeta$\textbf{
then}}

\STATE{~~~~~~~~~~~~\textbf{set} $\boldsymbol{g}\leftarrow\mathbf{0}$}

\STATE{~~~~~~~~~~~~\textbf{set} $cont\leftarrow false$}

\STATE{~~~~~~~~~\textbf{else}}

\STATE{~~~~~~~~~~~~\textbf{if} $a_{i}>\zeta$\textbf{
then}}

\STATE{~~~~~~~~~~~~~~~~\textbf{set} $z_{i}\leftarrow1$}

\STATE{~~~~~~~~~~~~\textbf{else}}

\STATE{~~~~~~~~~~~~~~~~\textbf{set} $z_{i}\leftarrow-1$}

\STATE{~~~~\textbf{if} $cont=true$ \textbf{then}}

\STATE{~~~~~~~~\textbf{set} $\boldsymbol{g}\leftarrow\mathbf{\mathbf{S}^{\dagger}}\boldsymbol{z}$}

\STATE{~~~~\textbf{if} $\left\Vert \boldsymbol{g}\right\Vert _{2}>\zeta$
\textbf{then}}

\STATE{~~~~~~~\textbf{set} $\boldsymbol{d}\leftarrow-\left(\boldsymbol{g}/\left\Vert \boldsymbol{g}\right\Vert _{2}\right)$}

\STATE{\textbf{end} \textbf{procedure}}

\end{algorithmic}

The Shearlet-Based Superiorized Version of ART makes use of the above
procedure by calling \textbf{\textsc{NonascendingShearlet$\left(\boldsymbol{x}^{\left(k,n\right)},\boldsymbol{v}^{\left(k,n\right)},\zeta\right)$}}
in Step \ref{Line:NAV} of Algorithm \ref{alg:Superiorized_ART}.
Accordingly, in addition to the already listed user-specified input
parameters for the \textbf{Superiorized Version of ART}, the Shearlet-Based
Superiorized algorithm requires user-specification of $\zeta$. (Recall
that the same is true for the TV-Based Superiorized algorithm.)

We emphasize once more that the superiorization approach just described
does not depend on $\mathbf{S}$ being the matrix associated with
the discrete shearlet transform, and so it is applicable to any transform
that can be specified by any matrix $\mathbf{S}.$ 
\begin{figure}[t]
\begin{centering}
\begin{tabular}{cc}
\subfloat[]{\label{subfig:Head_Phantom}\includegraphics[scale=0.18]{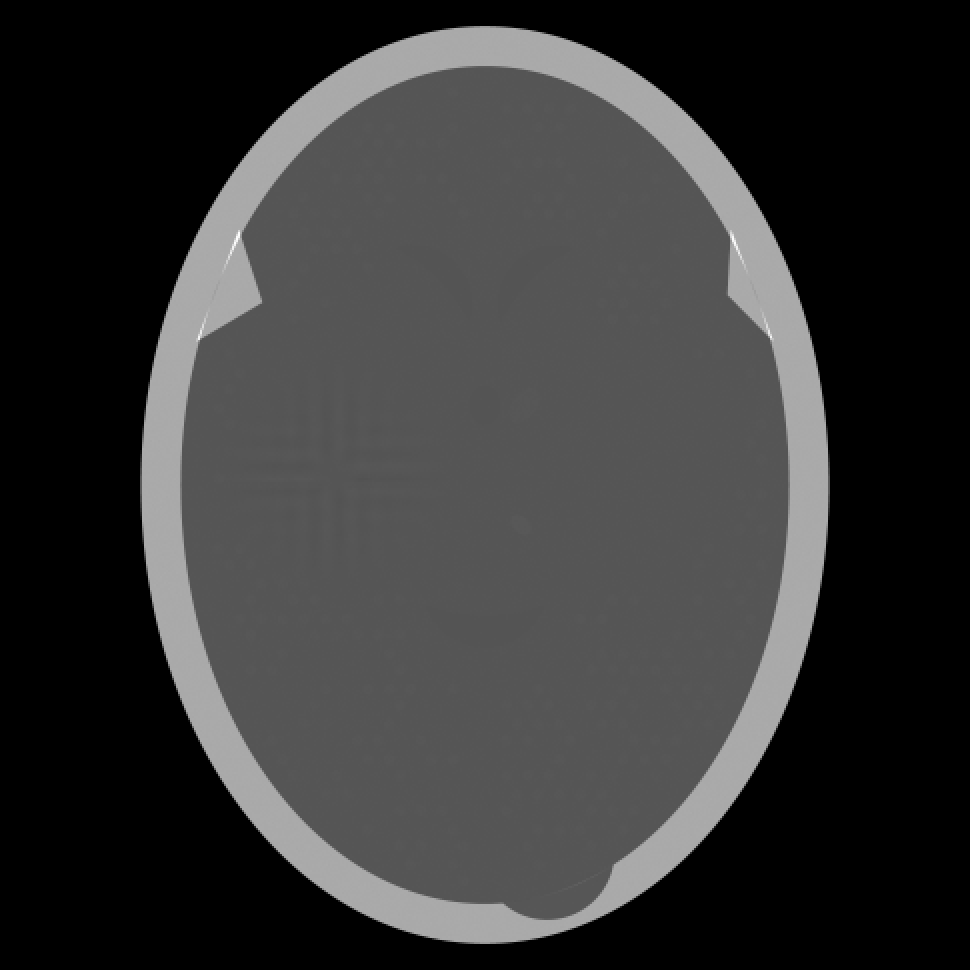}} & \subfloat[]{\label{subfig:Phantom_Thresholded}\includegraphics[scale=0.18]{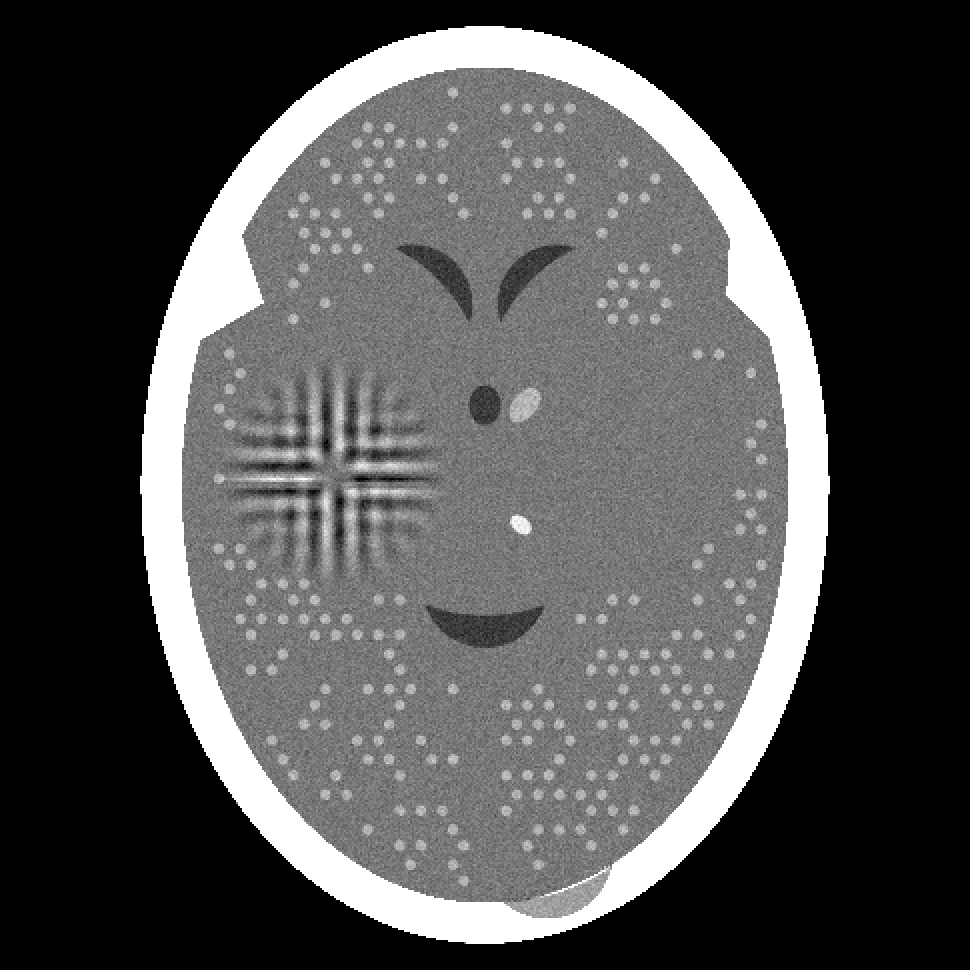}}\tabularnewline
\end{tabular}
\par\end{centering}
\centering{}\caption{\label{fig:Head_Phantom}A $485\times485$ digitization of a phantom
based on the distribution of x-ray attenuation in units of cm\protect\textsuperscript{-1}
within a transaxial slice of the human head. We display in (a) 0.000
as black and any value that is 0.6241749 or greater as white, in (b)
0.20400 as black and any value that is 0.21675 or greater as white.
The latter mapping of values into displayed intensities is used for
all the images shown below.}
\end{figure}

\section{{\normalsize{}Experiments and Analysis\label{sec:Experiments}}}

In this section we report on experiments with phantoms based on a
transaxial slice of the human head. The phantoms mimic an actual medical
image; for details, see Sections 4.3 and 4.4 of \cite{HERMAN:2009a}.
The various phantoms differ from each other by the random assignment
of local inhomogeneities and, more importantly, by a random introduction
of small ``tumors''; see Section 5.2 of \cite{HERMAN:2009a}. A
$485\times485$ digitization of one such phantom (produced by the
software SNARK14 \cite{SNARK14} in the manner specified in \cite{HERMAN:2009a})
is shown in Fig. \ref{fig:Head_Phantom}. In this, and in all of the
other digitized images shown in this paper, the length of a side of
a pixel is 0.0376 cm.

\subsection{{\normalsize{}Comparison Using Single Data Sets\label{subsec:Experiments_Single-Comparison}}}

We first describe an anecdotal experiment that compares outputs of
the classical methods of filtered back-projection (FBP) \cite[Chapter 10]{HERMAN:2009a}
and the algorithm ART with the reconstruction algorithms that are
discussed in the previous section; namely, the TV-Based Superiorized
Version of ART and the Shearlet-Based Superiorized Version of ART.
To make these comparisons, we applied the algorithms to CT problems
using the phantom of Fig. \ref{fig:Head_Phantom}. In the simulated
CT scanner, the acquisition process generated divergent projection
data, with source-to-detector distance 110.735 cm and source-to-center-of-rotation
distance 78 cm, over view angles with 693 rays per view, with a detector
spacing of 0.0533 cm. The projection data were simulated using integrals
over the original structures rather than over digitized versions of
them. The stochastic nature of the data collection is simulated by
using 1,000,000 photons for estimating each line integral. The phantom
and data acquisition were generated using the software SNARK14 \cite{SNARK14}.
The SNARK14 software allows the modeling of beam hardening that would
be experienced in a real CT scanner (for exact details, see the description
of the \emph{standard projection data} in Section 5.8 of \cite{HERMAN:2009a}),
but here we did not make use of this feature. The SNARK14 software
was also used for implementing the various reconstruction algorithms
in the experiments.

\begin{figure}
\noindent \begin{centering}
\begin{tabular}{cc}
\subfloat[]{\label{fig:FBP_0180}\includegraphics[bb=0bp 0cm 970bp 970bp,scale=0.18]{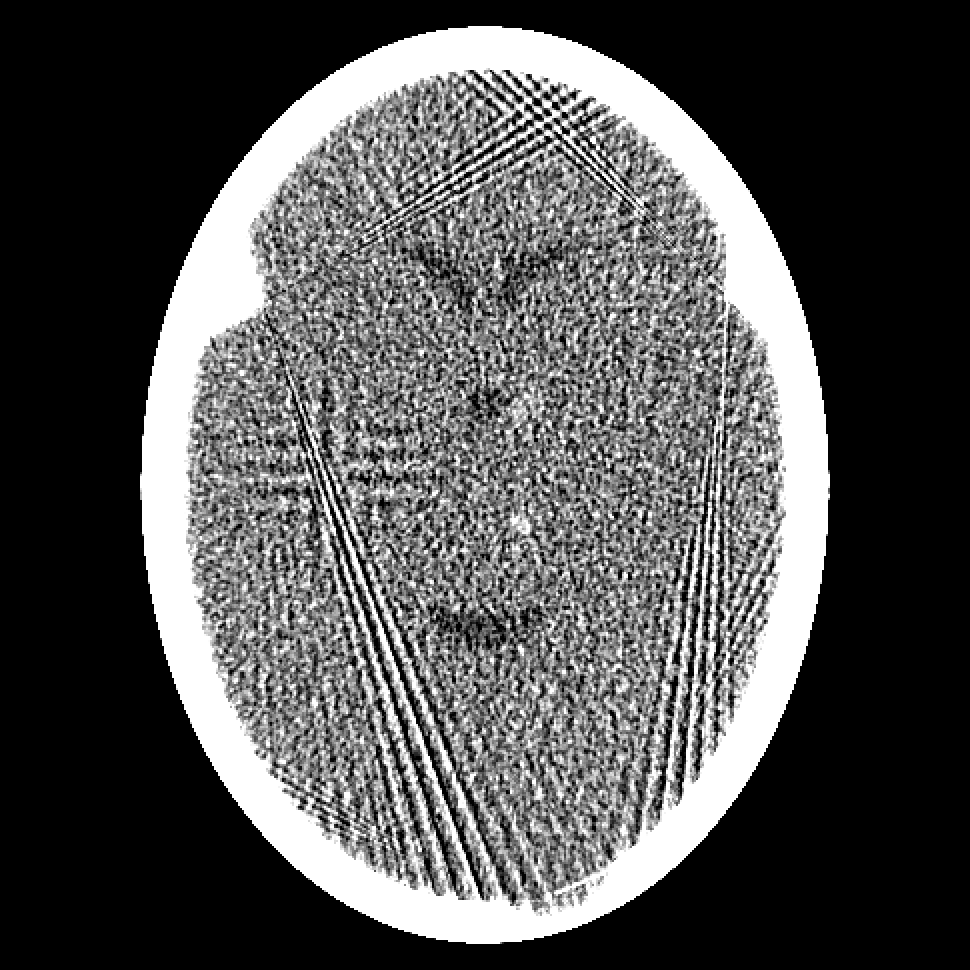}} & \subfloat[]{\label{fig:ARTP_0180}\includegraphics[bb=0bp 0bp 970bp 970bp,scale=0.18]{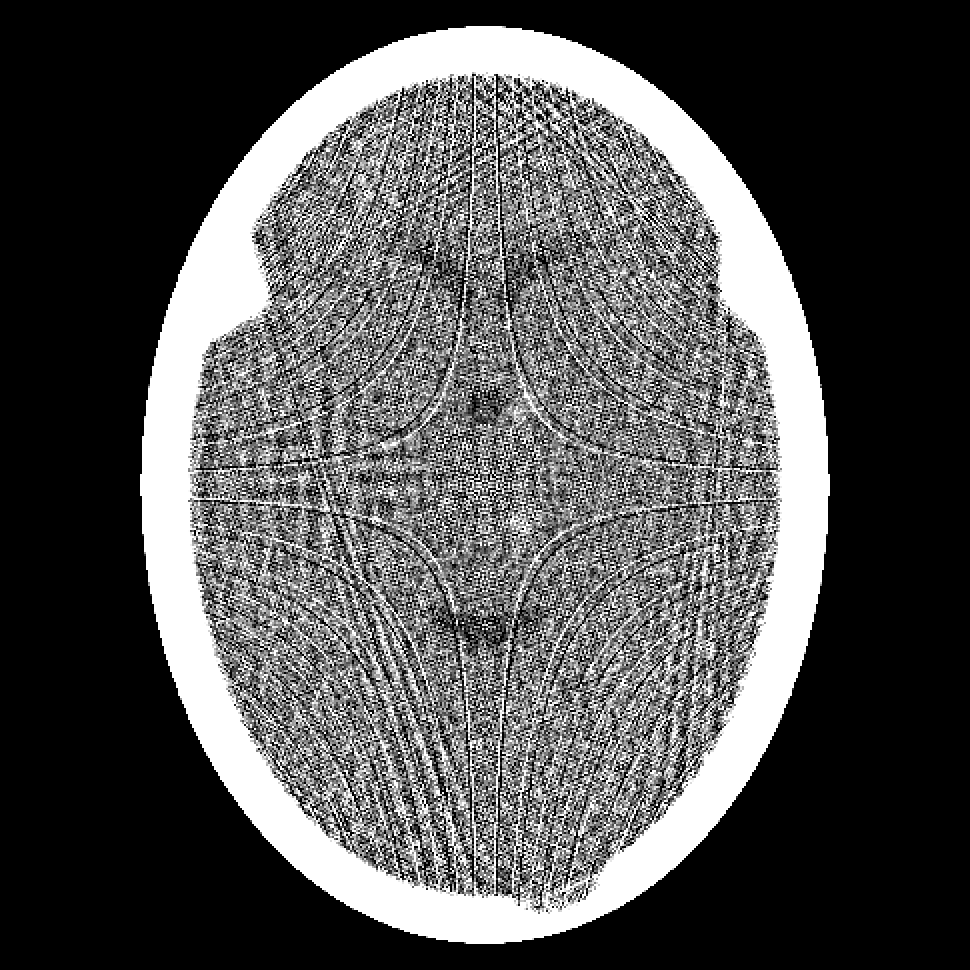}}\tabularnewline
\subfloat[]{\label{fig:ARTV_0180}\includegraphics[scale=0.18]{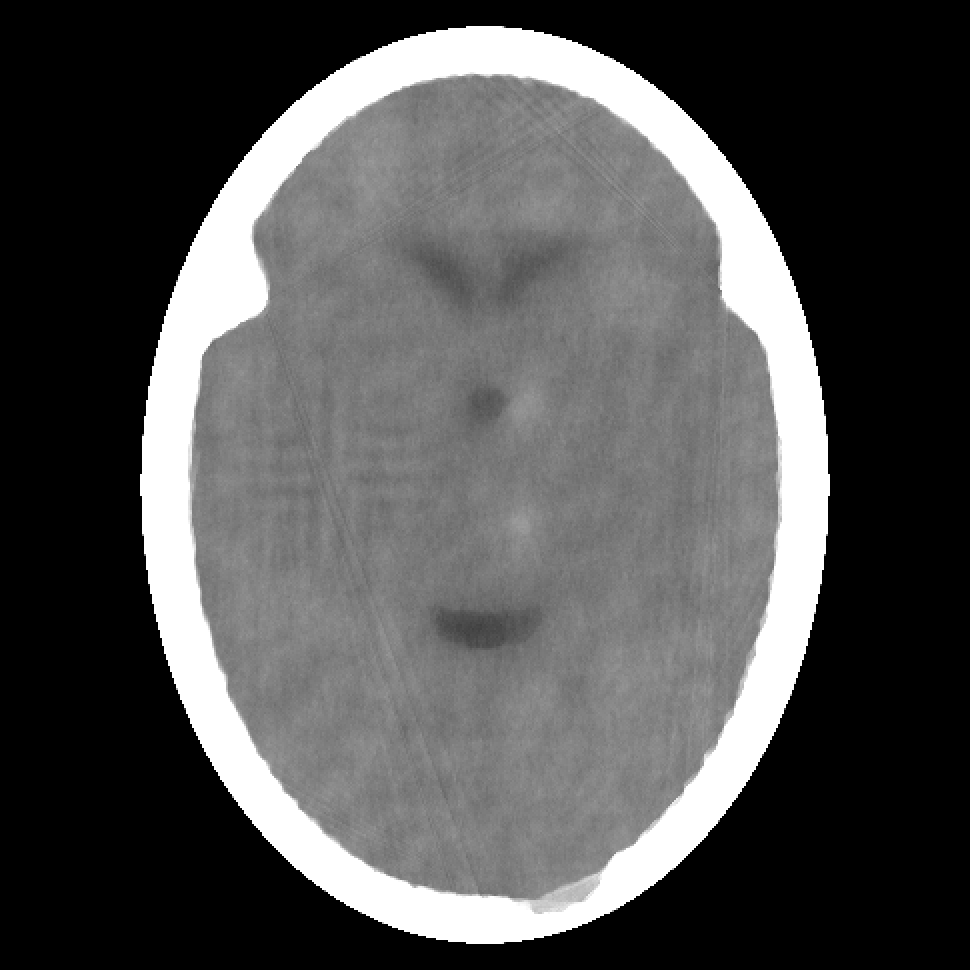}} & \subfloat[]{\label{fig:ARSH_0180}\includegraphics[scale=0.18]{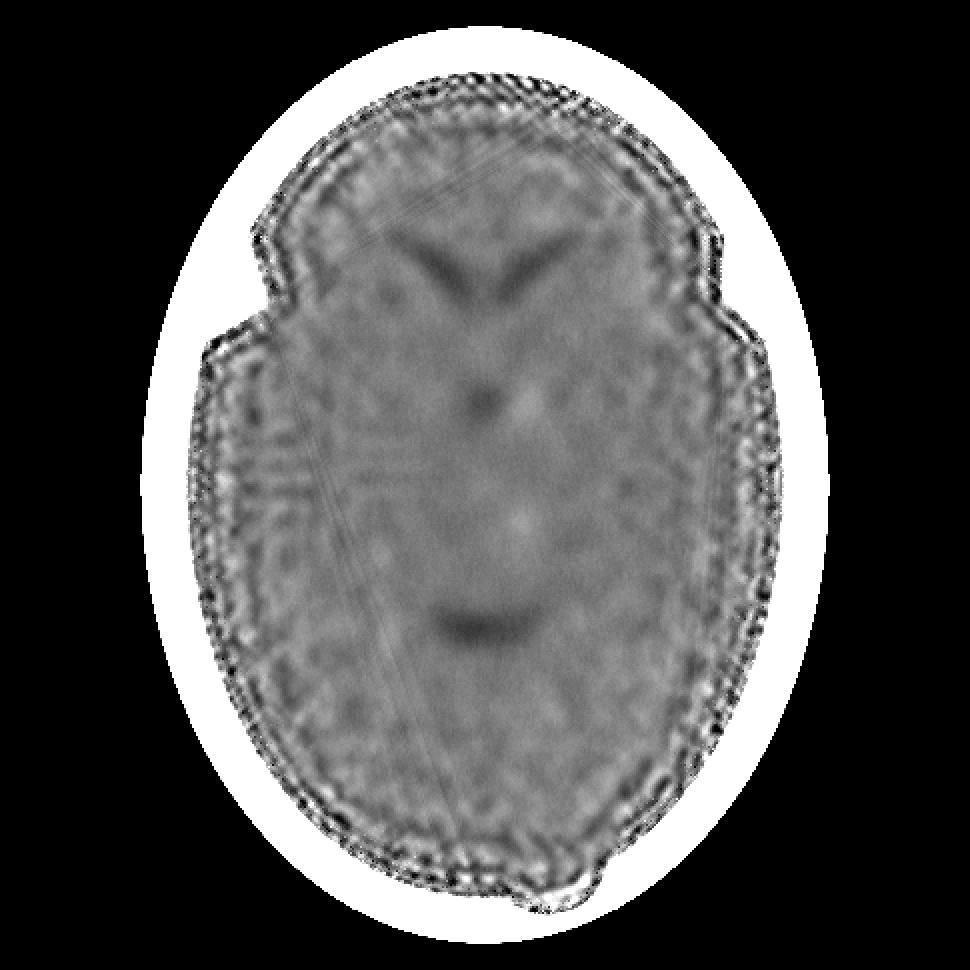}}\tabularnewline
\end{tabular}
\par\end{centering}
\centering{}\caption{\label{fig:Results_0180}Reconstructions from 180 projections by (a)
filtered back-projection, (b) the algorithm ART, (c) TV-Based Superiorized
Version of ART, and (d) Shearlet-Based Superiorized Version of ART.}
\end{figure}

\begin{figure}
\noindent \begin{raggedright}
\begin{tabular}{cc}
\subfloat[]{\label{fig:FBP_0360}\includegraphics[scale=0.18]{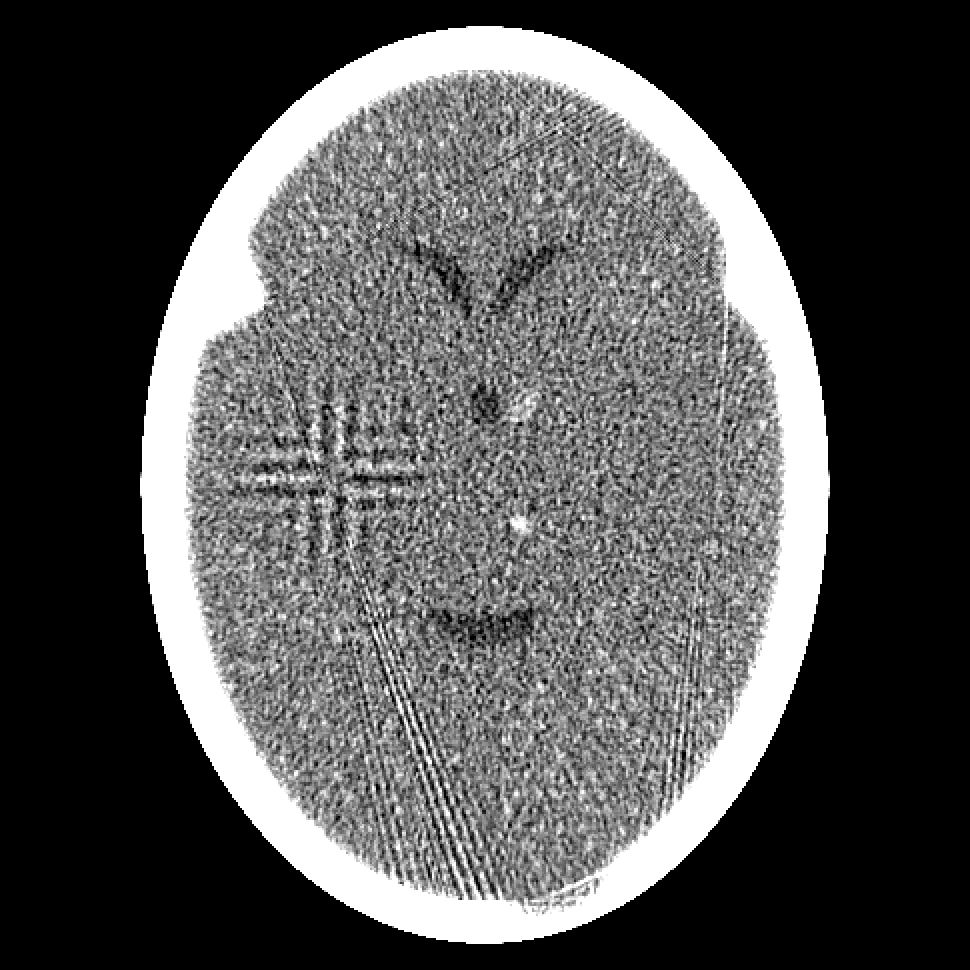}} & \subfloat[]{\label{fig:ARTP_0360}\includegraphics[scale=0.18]{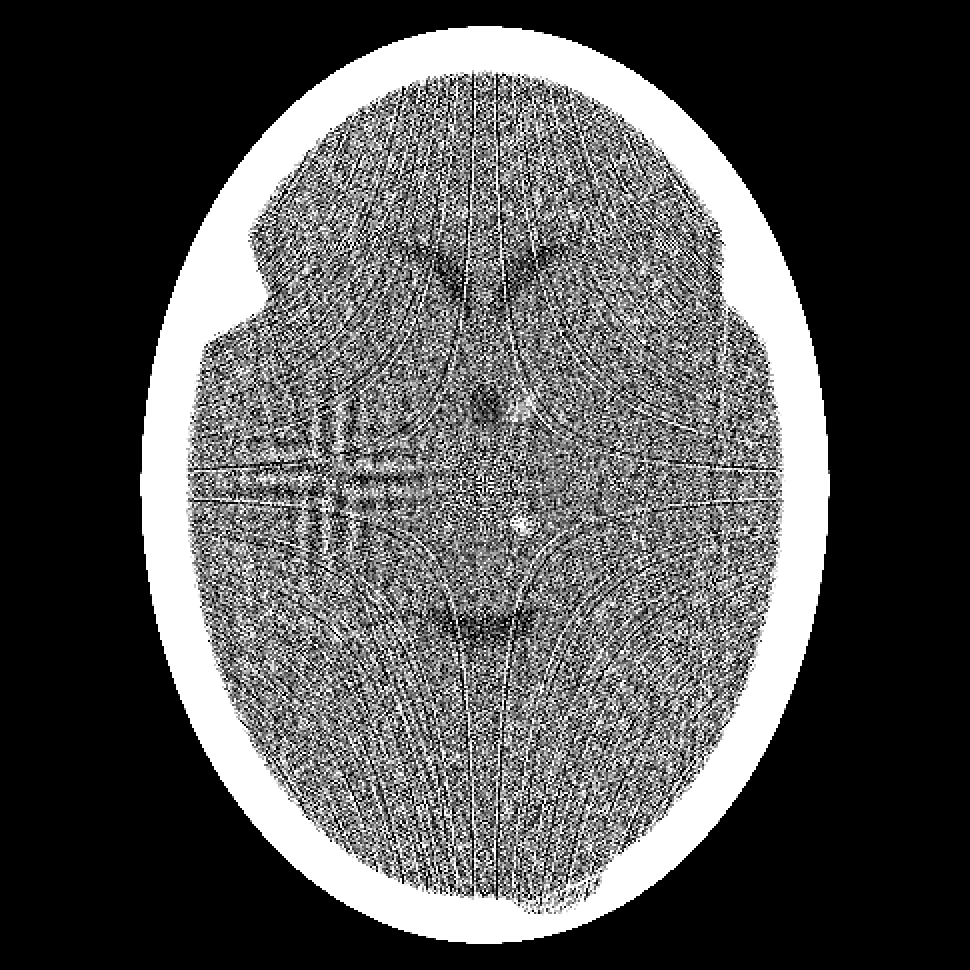}}\tabularnewline
\subfloat[]{\label{fig:ARTV_0360}\includegraphics[scale=0.18]{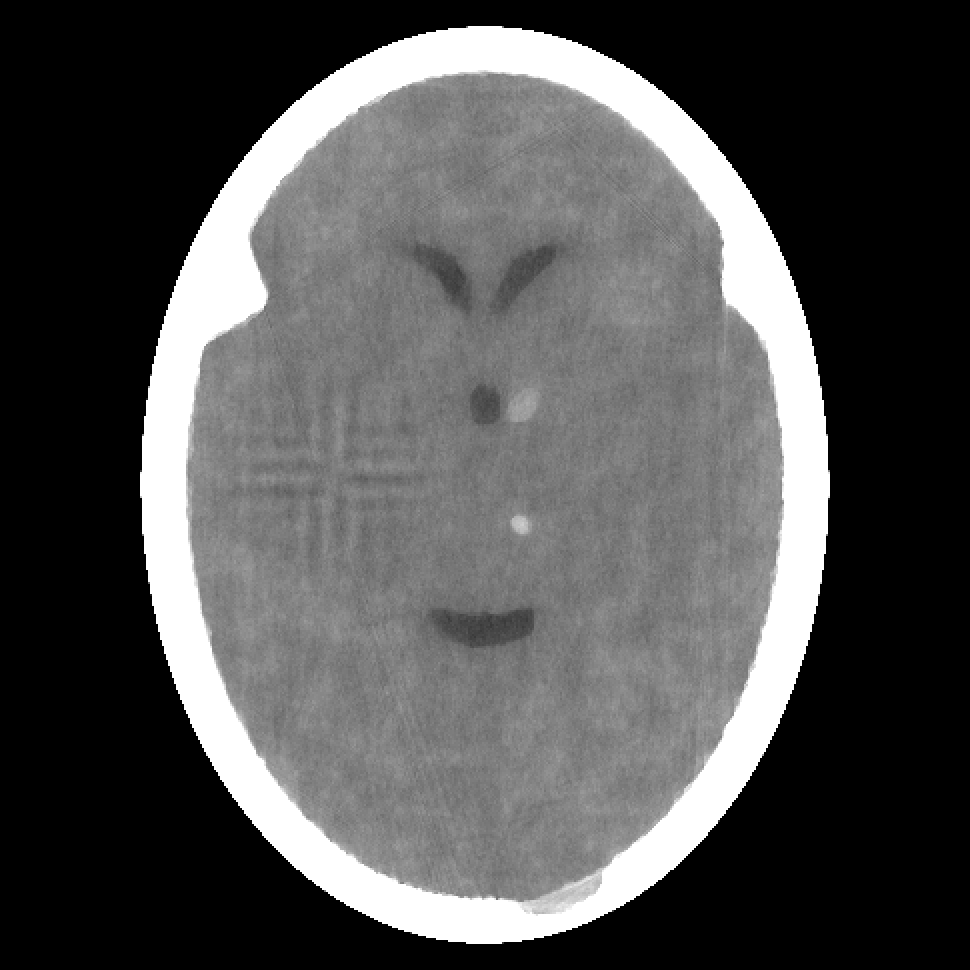}} & \subfloat[]{\label{fig:ARSH_0360}\includegraphics[scale=0.18]{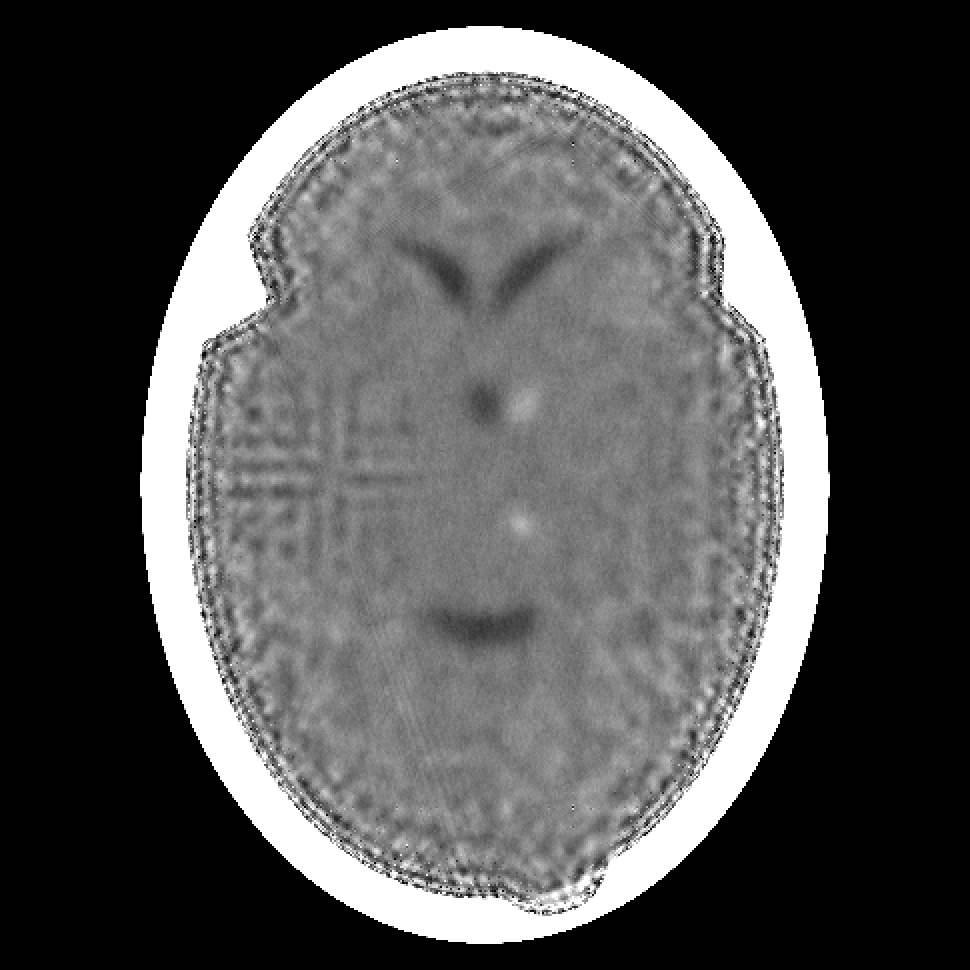}}\tabularnewline
\end{tabular}
\par\end{raggedright}
\centering{}\caption{\label{fig:Results_0360}Reconstructions from 360 projections by (a)
filtered back-projection, (b) the algorithm ART, (c) TV-Based Superiorized
Version of ART, and (d) Shearlet-Based Superiorized Version of ART.}
\end{figure}

\begin{figure}
\noindent \begin{centering}
\begin{tabular}{cc}
\subfloat[]{\label{fig:FBP_0720}\includegraphics[scale=0.18]{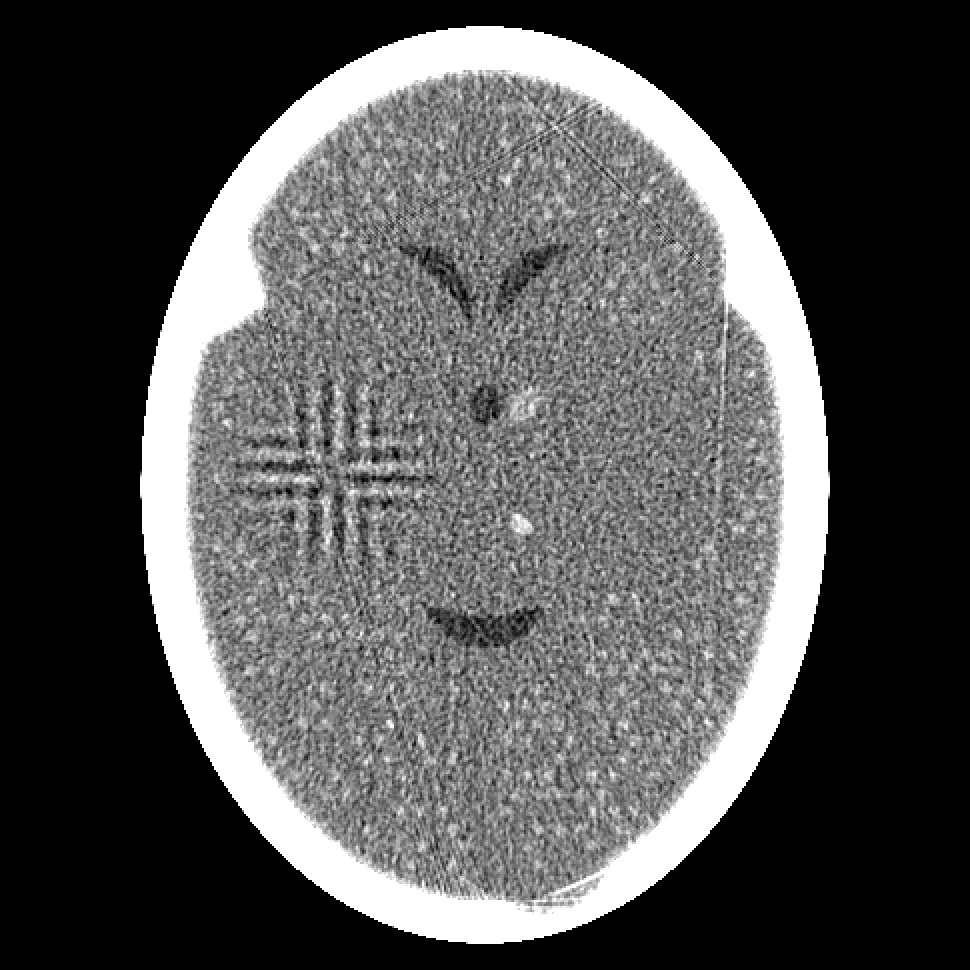}} & \subfloat[]{\label{fig:ARTP_0720}\includegraphics[scale=0.18]{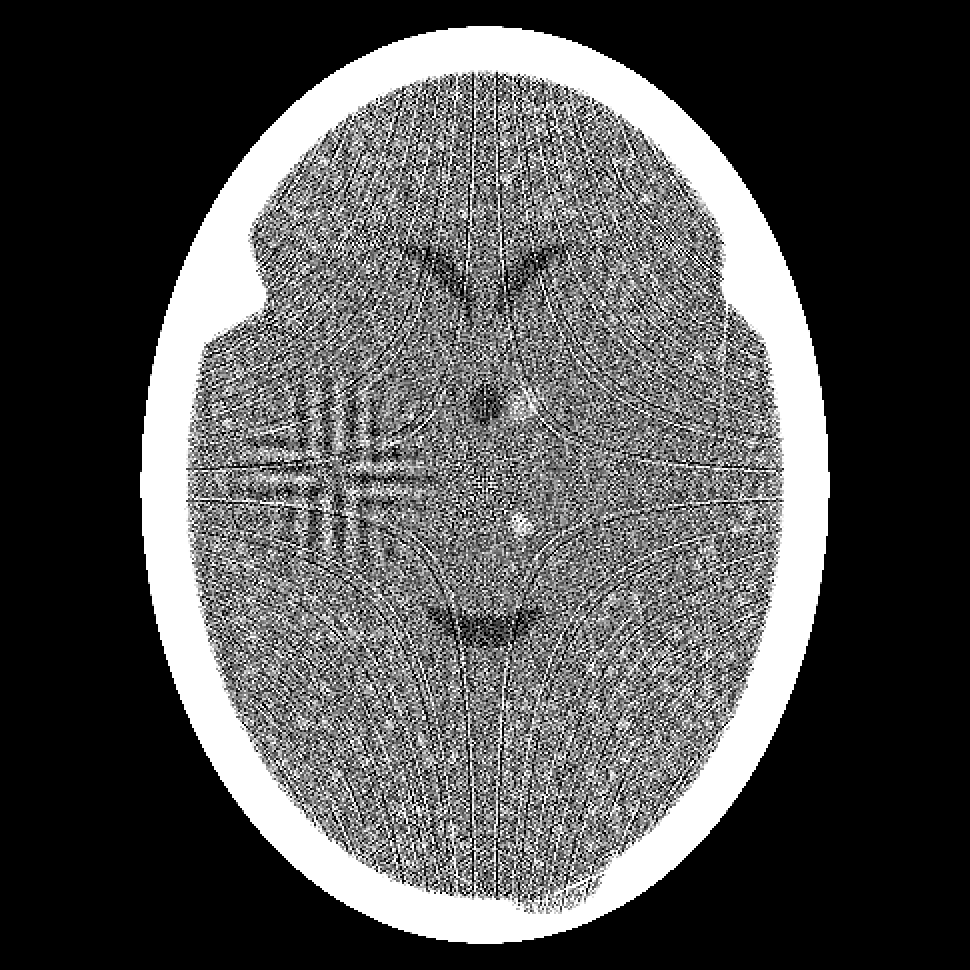}}\tabularnewline
\subfloat[]{\label{fig:ARTV_0720}\includegraphics[scale=0.18]{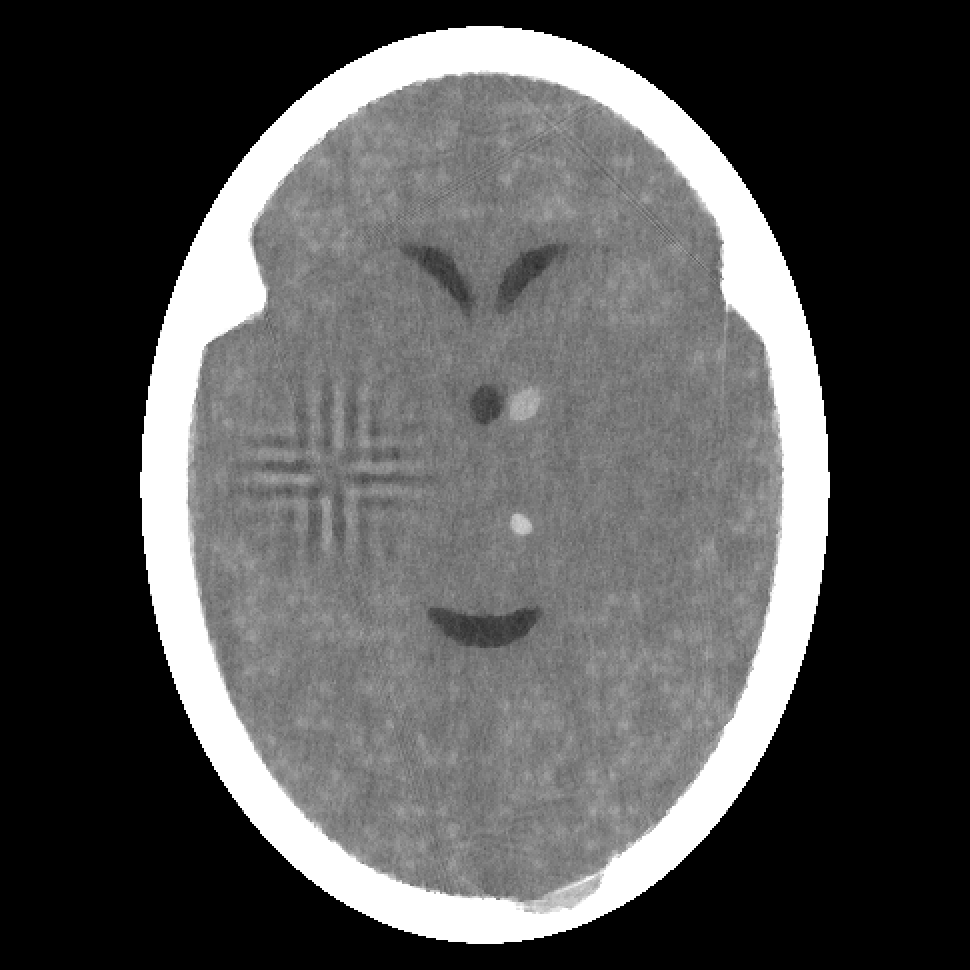}} & \subfloat[]{\label{fig:ARSH_0720}\includegraphics[scale=0.18]{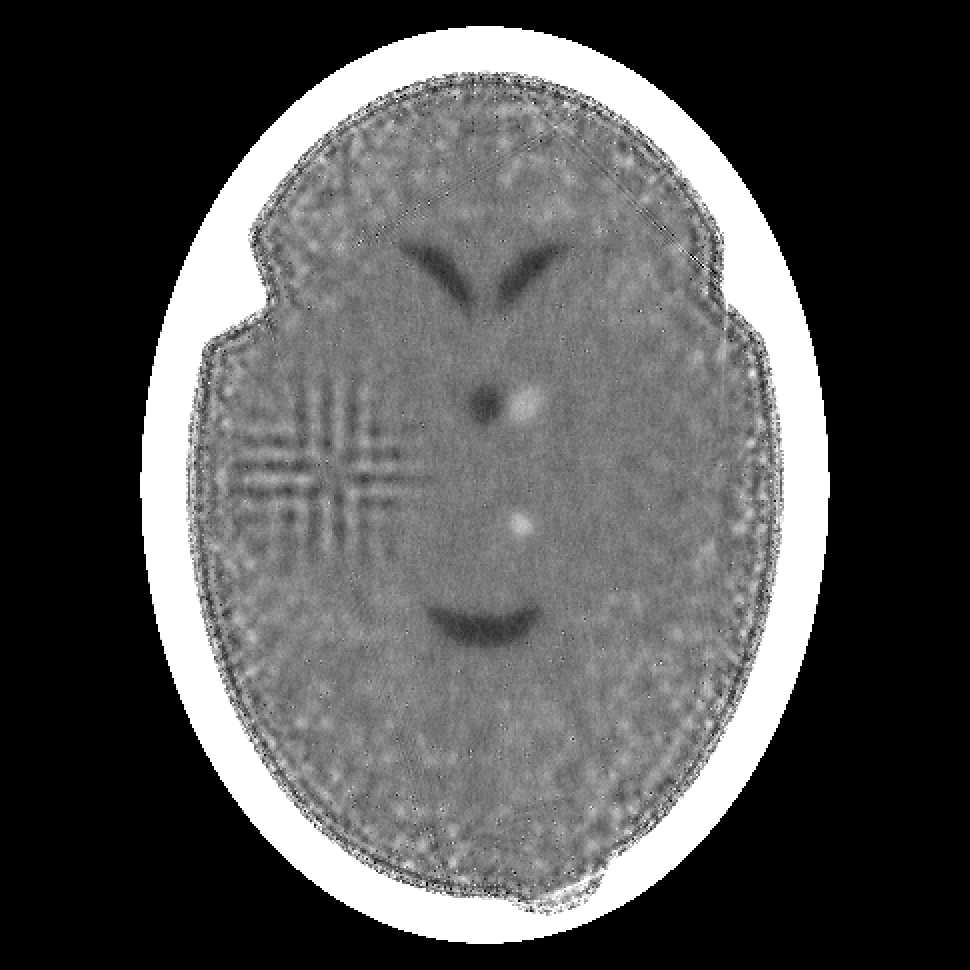}}\tabularnewline
\end{tabular}
\par\end{centering}
\centering{}\caption{\label{fig:Results_0720}Reconstructions from 720 projections by (a)
filtered back-projection, (b) the algorithm ART, (c) TV-Based Superiorized
Version of ART, and (d) Shearlet-Based Superiorized Version of ART.}
\end{figure}

For this anecdotal experiment we use three different numbers of views
(i.e., projections): 180, 360, and 720. We emphasize that, in the
currently-described anecdotal experiment, there is only one phantom
(which provides the ground truth); random generation of local inhomogeneities
and of tumor locations is done only once and the same arrangement
of local inhomogeneities and of tumor locations is used when generating
the projection data for the three different numbers of views.

The details of the reconstruction algorithms that we compare are as
follows. The specific choices are based on published results and some
preliminary experiments.
\begin{itemize}
\item For the filtered back-projection (FBP) method \cite[Chapter 10]{HERMAN:2009a},
we used the sinc window with linear interpolation (also called the
Shepp-Logan window).
\item Each of the three iterative algorithms returns as its output the vector
$\boldsymbol{x}^{(k)}$ for the smallest value of $k$ such that $\left\Vert \boldsymbol{y}-\mathbf{R}\boldsymbol{x}^{(k)}\right\Vert _{2}\leq\left\Vert \boldsymbol{y}-\mathbf{R}\boldsymbol{x}\right\Vert _{2}$,
where $\boldsymbol{x}$ is the output returned by FBP for the same
projection data $\boldsymbol{y}$.
\item For the two superiorized versions of ART, we used the values of 0.03
for $\beta_{0}$, 0.9999 for $\alpha$, 0.05 for $\rho$, 40 for $N$,
and $10^{-20}$ for $\zeta$.
\end{itemize}
\noindent We present the visual results of the reconstructions produced
by these algorithms when using 180, 360, and 720 projections in Figures
\ref{fig:Results_0180}, \ref{fig:Results_0360}, and \ref{fig:Results_0720},
respectively. We now give our impressions based on these visual results.

From the results for the single data set with 180 projections (Figure
\ref{fig:Results_0180}), we see that none of the four reconstruction
algorithms produces an image in which the small tumors are easily
locatable. Furthermore, the Shearlet-Based Superiorized Version of
ART introduces high frequency artifacts in the brain near the skull
and blurs the features inside the brain. In comparison, both filtered
back-projection and ART (to a lesser extent) introduce artifacts in
the form of streaks originating from interior bone edges. The image
produced by TV-superiorized ART does not show significant high-frequency
artifacts, the only one of the four, but the image is blurred.

In the case of 360 projections, the resulting images still show that
none of the algorithms provides clear visualization of the small tumors;
although they are somewhat visible in the images produced by filtered
back-projection and by ART. The Shearlet-Based Superiorized Version
of ART still shows high-frequency artifacts, albeit less pronounced
than in the case of 180 projections. TV-superiorized ART produces
an image in which the features, especially the small tumors, are smoothed
out, but the larger features within the brain are clearly identifiable.

As expected, the greater are the number of projections, the better
are the reconstructions produced by any of the four algorithms. However,
even with 720 projections, the reconstruction produced by the Shearlet-Based
Superiorized Version of ART still shows some high-frequency artifacts
and the small tumors are less visible than in the images produced
by the other three algorithms.

To supplement the just-listed visual impressions with numerical results,
we computed the residual (i.e., $\left\Vert \boldsymbol{y}-\mathbf{R}\boldsymbol{x}\right\Vert _{2}$),
the total variation (i.e., $\Upsilon\left(\boldsymbol{x}\right)$),
and the $\ell_{1}$-norm of the shearlet transform (i.e., $\left\Vert \mathbf{S}\boldsymbol{x}\right\Vert _{1}$)
for all the reconstructions produced by the algorithms when using
180, 360, and 720 projections; we present these values in Table \ref{tab:values_TV_SH}.
The entries in the table indicate that the presented superiorized
versions manifest what is expected from them: For all three data sets,
the value of $\Upsilon\left(\boldsymbol{x}\right)$ for the $\boldsymbol{x}$
produced by the TV-Based Superiorized Version of ART is smaller than
the value of $\Upsilon\left(\boldsymbol{x}\right)$ for the $\boldsymbol{x}$
produced by any of the other three algorithms and the value of $\left\Vert \mathbf{S}\boldsymbol{x}\right\Vert _{1}$
for the $\boldsymbol{x}$ produced by the Shearlet-Based Superiorized
Version of ART is smaller than the value of $\left\Vert \mathbf{S}\boldsymbol{x}\right\Vert _{1}$
for the $\boldsymbol{x}$ produced by any of the other three algorithms.
\begin{table}
\begin{tabular}{|c|c|c|c|c|c|}
\hline 
\# of & \multirow{2}{*}{Measure} & \multirow{2}{*}{FBP} & \multirow{2}{*}{ART} & Superiorized & Superiorized\tabularnewline
Angles &  &  &  & ART $\Upsilon\left(\boldsymbol{x}\right)$ & ART $\left\Vert \mathbf{S}\boldsymbol{x}^{\left(k\right)}\right\Vert _{1}$\tabularnewline
\hline 
\hline 
\multirow{3}{*}{180} & $\left\Vert \boldsymbol{y}-\mathbf{R}\boldsymbol{x}\right\Vert _{2}$ & 3.6380 & 3.6088 & 3.5389 & 3.6235\tabularnewline
\cline{2-6} 
 & $\Upsilon\left(\boldsymbol{x}\right)$ & 3,007.6751 & 3,565.0785 & 926.5716 & 1,261.0999\tabularnewline
\cline{2-6} 
 & $\left\Vert \mathbf{S}\boldsymbol{x}\right\Vert _{1}$ & 7,289.0860 & 6,779.7905 & 4,935.8958 & 4,673.0563\tabularnewline
\hline 
\hline 
\multirow{3}{*}{360} & $\left\Vert \boldsymbol{y}-\mathbf{R}\boldsymbol{x}\right\Vert _{2}$ & 4.0314 & 3.9267 & 3.9089 & 4.0075\tabularnewline
\cline{2-6} 
 & $\Upsilon\left(\boldsymbol{x}\right)$ & 1,797.8089 & 3,259.0070 & 955.4895 & 1,268.0752\tabularnewline
\cline{2-6} 
 & $\left\Vert \mathbf{S}\boldsymbol{x}\right\Vert _{1}$ & 5,840.9464 & 6,672.6732 & 5,278.3972 & 5,031.3871\tabularnewline
\hline 
\hline 
\multirow{3}{*}{720} & $\left\Vert \boldsymbol{y}-\mathbf{R}\boldsymbol{x}\right\Vert _{2}$ & 5.9793 & 5.3769 & 5.7150 & 5.7747\tabularnewline
\cline{2-6} 
 & $\Upsilon\left(\boldsymbol{x}\right)$ & 1,331.3471 & 2,900.1962 & 1,016.4190 & 1,346.1710\tabularnewline
\cline{2-6} 
 & $\left\Vert \mathbf{S}\boldsymbol{x}\right\Vert _{1}$ & 5,389.4274 & 6,444.8295 & 5,227.6876 & 5,147.9080\tabularnewline
\hline 
\end{tabular}

\caption{\label{tab:values_TV_SH}Computed values for the residual (i.e., $\left\Vert \boldsymbol{y}-\mathbf{R}\boldsymbol{x}\right\Vert _{2}$),
the total variation (i.e., $\Upsilon\left(\boldsymbol{x}\right)$),
and the $\ell_{1}$-norm of the shearlet transform (i.e., $\left\Vert \mathbf{S}\boldsymbol{x}\right\Vert _{1}$)
for all the reconstructions produced by filtered back-projection (FBP),
the algorithm ART, the TV-Based Superiorized Version of ART (using
$\Upsilon\left(\boldsymbol{x}\right)$), and the Shearlet-Based Superiorized
Version of ART (using $\left\Vert \mathbf{S}\boldsymbol{x}\right\Vert _{1})$,
for 180, 360, and 720 projection angles.}
\end{table}

That the mathematical behavior of the algorithms is as expected indicates
the correctness of the theory and the programming, but says nothing
about the relative efficacy in practice of using TV or shearlets for
the secondary criterion. We now turn to discussing this topic.

First we analyze the anecdotal experiment from this point of view.
One figure of merit for the efficacy of a reconstruction algorithm
is the \textit{relative error}, which is a normalized mean absolute
distance measure \cite[equation (5.2)]{HERMAN:2009a}. In our current
notation it is defined as
\begin{equation}
r=\left\Vert \mathbf{p}-\boldsymbol{x}\right\Vert _{1}/\left\Vert \mathbf{p}\right\Vert _{1},\label{eq:relative_error}
\end{equation}
where $\boldsymbol{p}$ is the phantom and $\boldsymbol{x}$ is the
reconstruction.

\begin{figure}
\centering\subfloat[]{\includegraphics[scale=0.39]{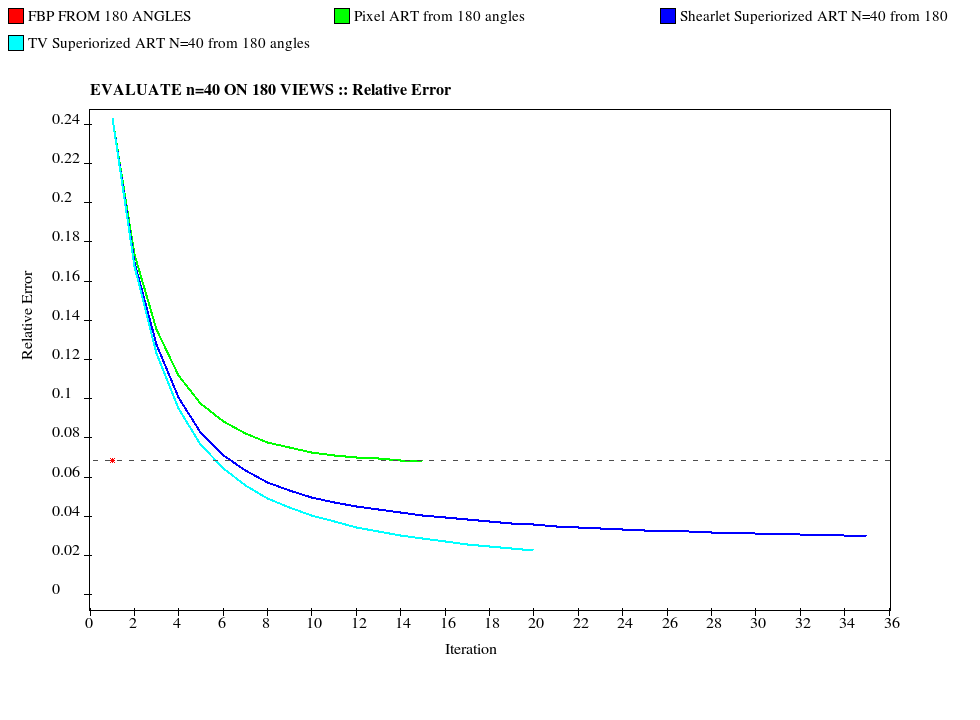}}

\subfloat[]{\includegraphics[scale=0.39]{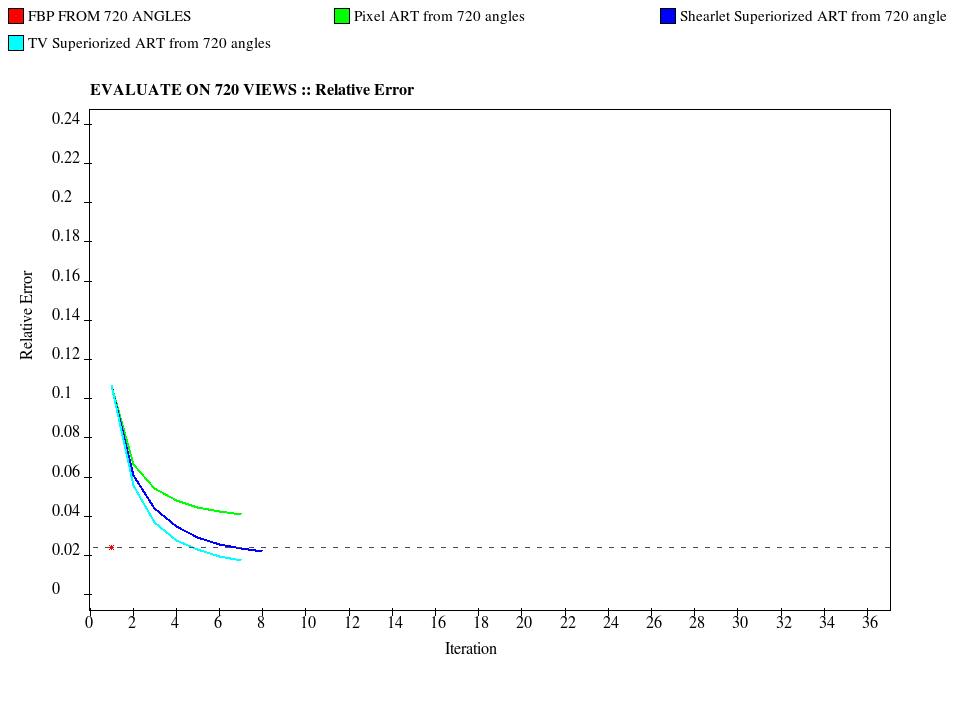}}\caption{\label{fig:The-relative-errors}The relative errors (\ref{eq:relative_error})
of the reconstructions produced by the four algorithms for two data
sets: (a) 180 projections and (b) 720 projections.}
\end{figure}

In Fig. \ref{fig:The-relative-errors} we report on the relative errors
for the data sets of 180 and 720 views. In both cases the red dot
(with the dashed line) indicates the value of the relative error obtained
by FBP (a noniterative algorithm). For the iterative algorithms, we
report on the relative error for $\boldsymbol{x}=\boldsymbol{x}^{(k)}$,
where $k$ is the iteration number. We see that, for both data sets,
the TV-Based Superiorized Version of ART outperforms the Shearlet-Based
Superiorized Version of ART, at least when performance is measured
by the relative error $r$. Both superiorized versions outperform
the unsuperiorized algorithm ART. The same is true for the data set
for 360 views; we are not illustrating that in this paper.

\subsection{{\normalsize{}Statistical Comparison\label{subsec:SHT-Comparison}}}

In order to draw firm conclusions about the relative efficacy of the
four reconstruction algorithms, we complement our report based on
single data sets with other experiments that use \textit{statistical
hypothesis testing} (SHT) for task-oriented comparisons \cite[Section 5.2]{HERMAN:2009a}
of them. Similarly to the experiments for the single data sets, we
used three different number of projections (180, 360, 720) for the
SHT experiments. The reported results were obtained by SNARK14 using
its ``experimenter'' feature \cite[Chapter 8]{SNARK14}.

Our SHT experiments consist of the following four steps: (\emph{i})
Generation of samples from an ensemble of phantoms of the kind described
at the beginning of this Section \ref{sec:Experiments}. The ensemble
is based on a transaxial slice of the human head with local inhomogeneities.
Note that this by itself provides us a statistical ensemble because
the local inhomogeneities are introduced using a Gaussian random variable.
However there is an additional (for the task more relevant) variability
within the ensemble that is achieved as follows. We specify a large
number of pairs of potential tumor sites, the locations of the sites
in a pair are symmetrically placed in the left and right halves of
the brain. In any sample from the ensemble, exactly one of each pair
of the sites will actually have a tumor placed there, with equal probability
for either site. In Fig. \ref{fig:Head_Phantom}(b) we illustrate
one sample from this ensemble (i.e., one of the phantoms with random
allocation of the tumors to the potential sites). (\emph{ii}) Generation
of realistic projection data sets, by using the parameters specified
for the CT simulator (this introduces extra randomness due to the
quantum noise in the data generation) for every randomly-generated
phantom and using them to produce reconstructions with the reconstruction
algorithms. (\emph{iii}) Measuring the goodness of every reconstruction
by using a medically-relevant \emph{figure} \emph{of} \emph{merit}
(FOM). (\emph{iv}) Computation of the statistical significance (based
on the FOMs of all the reconstructions for all the data sets) by which
the null hypothesis that a pair of algorithms is equally good can
be rejected in favor of the alternative hypothesis that the reconstruction
algorithm with the higher average value of the FOM (call this Algorithm
1) is better than reconstruction algorithm with the lower average
value of the FOM (call this Algorithm 2).

In order to obtain statistically significant results, we sampled the
ensemble of phantoms and generated projection data 30 times. For each
of the thirty projection data sets, we subtracted from the FOM of
the reconstruction produced by Algorithm 1 the FOM of the reconstruction
produced by Algorithm 2. (Note that the null hypothesis of equal efficacy
of the two algorithms would imply that the expected value of this
difference is zero.) We define $s$ as the average of these differences
over the thirty data sets. There is a method (for details, see \cite[Section 5.2)]{HERMAN:2009a})
for estimating the so-called P-value, which is the probability under
the null hypothesis of obtaining a value for $s$ that is as large
or larger than what we actually obtained in the experiment. If the
null hypothesis were correct, we would not expect to come across an
$s$ for which the P-value is very small. Thus, the smallness of the
P-value is a measure of \emph{significance }for rejecting the null
hypothesis that the two reconstruction algorithms are equally good
according to our selected FOM in favor of the \textit{\emph{alternative
hypothesis}} that Algorithm 1 is better than Algorithm 2.

For all our SHT experiments we used the FOM called \emph{imagewise-region-of-interest}
(IROI) \cite{HERMAN:2009a,SNARK14,NARAYAN:1999a} that, for our experiments,
is defined as follows. In our phantoms there are pairs of potential
tumor locations in the brain; we index these pairs with $b=1,\ldots,B$.
For each pair, just one of the locations contains a tumor, that increases
the value of $f$ at that location. We define
\begin{equation}
{\small\begin{array}{ll}
\textrm{IROI}= & \left(\frac{\sum_{b=1}^{B}\left(\upsilon_{t}^{p}\left(b\right)-\upsilon_{n}^{p}\left(b\right)\right)}{\sqrt{\sum_{b=1}^{B}\left(\upsilon_{n}^{p}\left(b\right)-\frac{1}{B}\sum_{b'=1}^{B}\upsilon_{n}^{p}\left(b'\right)\right)^{2}}}\right)^{-1}\times\\
 & \left(\frac{\sum_{b=1}^{B}\left(\upsilon_{t}^{r}\left(b\right)-\upsilon_{n}^{r}\left(b\right)\right)}{\sqrt{\sum_{b=1}^{B}\left(\upsilon_{n}^{r}\left(b\right)-\frac{1}{B}\sum_{b'=1}^{B}\upsilon_{n}^{r}\left(b'\right)\right)^{2}}}\right),
\end{array}}\label{eq:IROIfom}
\end{equation}
where $\upsilon_{t}^{p}\left(p\right)$ (respectively, $\upsilon_{n}^{p}\left(p\right)$)
denotes the average density in the phantom of the structure of the
$b$th pair that is (respectively, is not) the tumor. Similarly, the
$\upsilon_{t}^{r}\left(p\right)$ (respectively, $\upsilon_{n}^{r}\left(p\right)$)
denotes the average density in the reconstruction of the structure
of the $b$th pair that is (respectively, is not) the tumor. If the
reconstruction is perfect, in the sense of being identical to the
phantom, then IROI=1. All the parameters, including the stopping criteria
(provided by the $\varepsilon$), for the reconstructions from each
of the 30 random data sets were the ones specified previously for
the experiment using a single data sets.

By using SHT for task-oriented comparisons of the four reconstruction
algorithms, with IROI as the figure of merit, we found that TV-superiorized
ART always outperformed ART and Shearlet-Based Superiorized Version
of ART with strong statistical significance (i.e., very small P-values),
see Table \ref{tab:SHT}. However, TV-superiorized ART only outperformed
filtered back-projection in the experiment with 180 projections; in
fact, in the other two experiments (i.e., 360 and 720 projections)
filtered back-projection outperforms the other three algorithms with
strong statistical significance, see Table \ref{tab:SHT}. The conclusion
that we can draw from this is that if the number of projections is
not very small (i.e., 360 or more) and the medical task is the localization
of very small tumors in the brain, then the secondary criterion of
having a small TV is not the appropriate one for turning the algorithm
ART into a superiorized version that outperforms FBP. This is because
the smoothing property of this secondary criterion results in the
very small tumors being blurred out in the reconstruction, as is illustrated
very clearly in Fig. \ref{fig:ARTV_0360}. (We note that having a
small $\left\Vert \mathbf{S}\boldsymbol{x}\right\Vert _{1}$ has proven
to be an even less appropriate secondary criterion for this medical
task.) On the positive side for TV as a secondary criterion, we note
that, with 180 projections, the TV-based superiorized ART significantly
outperforms the other three reconstruction algorithms. 

\begin{table*}
\noindent \begin{centering}
\begin{tabular}{|c|c|c|c|c|c|}
\hline 
\# of & \multirow{2}{*}{FBP} & \multirow{2}{*}{ART} & Superiorized & Superiorized ART & \multirow{2}{*}{P-value}\tabularnewline
Angles &  &  & ART $\Upsilon\left(\boldsymbol{x}\right)$ & ART $\left\Vert \mathbf{S}\boldsymbol{x}\right\Vert _{1}$ & \tabularnewline
\hline 
\hline 
\multirow{6}{*}{180} & 0.070656 & 0.064593 &  &  & ${\scriptstyle 1.686016\times10^{-6}}$\tabularnewline
\cline{2-6} 
 & 0.070656 &  & 0.088268 &  & ${\scriptstyle 2.418553\times10^{-7}}$\tabularnewline
\cline{2-6} 
 & 0.070656 &  &  & 0.057435 & ${\scriptstyle 3.450878\times10^{-7}}$\tabularnewline
\cline{2-6} 
 &  & 0.064593 & 0.088268 &  & ${\scriptstyle 5.374371\times10^{-8}}$\tabularnewline
\cline{2-6} 
 &  & 0.064593 &  & 0.057435 & ${\scriptstyle 1.844853\times10^{-6}}$\tabularnewline
\cline{2-6} 
 &  &  & 0.088268 & 0.057435 & ${\scriptstyle 3.812557\times10^{-8}}$\tabularnewline
\hline 
\hline 
\multirow{6}{*}{360} & 0.163389 & 0.126363 &  &  & ${\scriptstyle 3.799954\times10^{-8}}$\tabularnewline
\cline{2-6} 
 & 0.163389 &  & 0.152645 &  & ${\scriptstyle 1.702295\times10^{-5}}$\tabularnewline
\cline{2-6} 
 & 0.163389 &  &  & 0.103251 & ${\scriptstyle 3.425471\times10^{-8}}$\tabularnewline
\cline{2-6} 
 &  & 0.126363 & 0.152645 &  & ${\scriptstyle 6.196416\times10^{-8}}$\tabularnewline
\cline{2-6} 
 &  & 0.126363 &  & 0.103251 & ${\scriptstyle 8.772077\times10^{-8}}$\tabularnewline
\cline{2-6} 
 &  &  & 0.152645 & 0.103251 & ${\scriptstyle 3.464983\times10^{-8}}$\tabularnewline
\hline 
\hline 
\multirow{6}{*}{720} & 0.235774 & 0.167454 &  &  & ${\scriptstyle 2.967738\times10^{-8}}$\tabularnewline
\cline{2-6} 
 & 0.235774 &  & 0.215847 &  & ${\scriptstyle 4.350362\times10^{-7}}$\tabularnewline
\cline{2-6} 
 & 0.235774 &  &  & 0.180246 & ${\scriptstyle 4.141425\times10^{-8}}$\tabularnewline
\cline{2-6} 
 &  & 0.167454 & 0.215847 &  & ${\scriptstyle 2.983396\times10^{-8}}$\tabularnewline
\cline{2-6} 
 &  & 0.167454 &  & 0.180246 & ${\scriptstyle 8.518713\times10^{-6}}$\tabularnewline
\cline{2-6} 
 &  &  & 0.215847 & 0.180246 & ${\scriptstyle 8.105134\times10^{-8}}$\tabularnewline
\hline 
\end{tabular}
\par\end{centering}
\caption{\label{tab:SHT}Results of the statistical hypothesis testing (SHT)
experiments that compared reconstructions produced by filtered back-projection
(FBP), the algorithm ART, the TV-Based Superiorized Version of ART
(using $\Upsilon\left(\boldsymbol{x}\right)$), and the Shearlet-Based
Superiorized Version of ART (using $\left\Vert \mathbf{S}\boldsymbol{x}\right\Vert _{1})$,
with IROI (\ref{eq:IROIfom}) as the figure of merit. The entries
in the body of this table are the average values of IROI over the
30 random data sets.}
\end{table*}

\section{The Split Bregman Approach\label{sec:split_Bregman_Method}}

The authors of \cite{VANDEGHINSTE:2013a} use a modified split Bregman
method with both the shearlet transform and total variation as regularization
terms. In this section we discuss such approaches and compare their
outputs against the superiorization method.

The CT problem of (\ref{eq:General_Optimization}) can be converted
into the regularized global optimization problem 
\begin{equation}
\mathbf{Find}\;\boldsymbol{x}^{*}=\arg\underset{\boldsymbol{x}}{\min}\left(\frac{\kappa}{2}\left\Vert \boldsymbol{y}-\mathbf{R}\boldsymbol{x}\right\Vert _{2}^{2}+\phi\left(\boldsymbol{x}\right)\right),\label{eq:Regularized_Problem}
\end{equation}
where $\kappa$ is a positive-real-number parameter that controls
the relative importance between the constraints-compatibility and
the prior desirability; this parameter replaces the $\varepsilon$
of (\ref{eq:General_Optimization}). An alternative way of describing
the role of $\kappa$ is that it determines the contribution of the
regularization term to the total cost (lower value of $\kappa$ results
in larger contribution of the regularization term $\phi\left(\boldsymbol{x}\right)$).
Formulations of both kinds (i.e., the ones of equations (\ref{eq:General_Optimization})
and (\ref{eq:Regularized_Problem})) are listed in the beginning parts
of \cite{GOLDSTEIN:2009a}; see also \cite{ZIBULEVSKY:2010a}.

The split Bregman method is an iterative procedure that, in addition
to producing a sequence $\boldsymbol{x}^{\left(k\right)}$ of $J$-dimensional
vectors that are supposed to converge to $\boldsymbol{x}^{*}$ of
(\ref{eq:Regularized_Problem}), also produces two auxiliary sequences
$\boldsymbol{q}^{\left(k\right)}$ and $\boldsymbol{b}^{\left(k\right)}$
of $I$-dimensional vectors according to the following recurrence
rules. We set $\boldsymbol{x}^{\left(0\right)}$, $\boldsymbol{q}^{\left(0\right)}$
and $\boldsymbol{b}^{\left(0\right)}$ to be vectors (each of them
denoted by $\mathbf{0})$ all of whose components are zero. For all
nonnegative integers $k$, we define
\begin{equation}
\boldsymbol{x}^{\left(k+1\right)}=\arg\underset{\boldsymbol{x}}{\min}\left(\frac{\kappa}{2}\left\Vert \boldsymbol{y}-\mathbf{R}\boldsymbol{x}\right\Vert _{2}^{2}+\frac{\mu}{2}\left\Vert \boldsymbol{q}^{\left(k\right)}-\phi\left(\boldsymbol{x}^{(k)}\right)-\boldsymbol{b}^{\left(k\right)}\right\Vert _{2}^{2}\right),\label{eq:Split-Bregman_a}
\end{equation}
where $\mu$ is a fixed positive-real-number relaxation parameter,
\begin{equation}
\boldsymbol{q}^{\left(k+1\right)}=\arg\underset{\boldsymbol{q}}{\min}\left(\left\Vert \boldsymbol{q}\right\Vert _{1}+\frac{\mu}{2}\left\Vert \boldsymbol{q}-\phi\left(\boldsymbol{x}^{(k+1)}\right)-\boldsymbol{b}^{\left(k\right)}\right\Vert _{2}^{2}\right),\label{eq:Split-Bregman_b}
\end{equation}
 
\begin{equation}
\boldsymbol{b}^{\left(k+1\right)}=\boldsymbol{b}^{\left(k\right)}+\left(\phi\left(\boldsymbol{x}^{(k+1)}\right)-\boldsymbol{q}^{\left(k+1\right)}\right).\label{eq:Split-Bregman_c}
\end{equation}

\subsection{Using the $\ell_{1}$-Norm of the Shearlet Transform}

The authors of \cite{VANDEGHINSTE:2013a} propose replacing $\phi\left(\boldsymbol{x}\right)$
by $\left\Vert \mathbf{S}\boldsymbol{x}\right\Vert _{1}$ in (\ref{eq:Regularized_Problem}),
resulting in equations (\ref{eq:Split-Bregman_a}), (\ref{eq:Split-Bregman_b})
and (\ref{eq:Split-Bregman_c}) becoming
\begin{equation}
\boldsymbol{x}^{\left(k+1\right)}=\arg\underset{\boldsymbol{x}}{\min}\left(\frac{\kappa}{2}\left\Vert \boldsymbol{y}-\mathbf{R}\boldsymbol{x}\right\Vert _{2}^{2}+\frac{\mu}{2}\left\Vert \boldsymbol{q}^{\left(k\right)}-\mathbf{S}\boldsymbol{x}^{(k)}-\boldsymbol{b}^{\left(k\right)}\right\Vert _{2}^{2}\right),\label{eq:Split-Bregman_SH_a}
\end{equation}
\begin{equation}
\boldsymbol{q}^{\left(k+1\right)}=\arg\underset{\boldsymbol{q}}{\min}\left(\left\Vert \boldsymbol{q}\right\Vert _{1}+\frac{\mu}{2}\left\Vert \boldsymbol{q}-\mathbf{S}\boldsymbol{x}^{(k+1)}-\boldsymbol{b}^{\left(k\right)}\right\Vert _{2}^{2}\right),\label{eq:Split-Bregman_SH_b}
\end{equation}
 
\begin{equation}
\boldsymbol{b}^{\left(k+1\right)}=\boldsymbol{b}^{\left(k\right)}+\left(\mathbf{S}\boldsymbol{x}^{(k+1)}-\boldsymbol{q}^{\left(k+1\right)}\right),\label{eq:Split-Bregman_SH_c}
\end{equation}
respectively. 

By taking the derivatives of the right hand side of (\ref{eq:Split-Bregman_SH_a}),
we see that the $\boldsymbol{x}^{\left(k+1\right)}$ of (\ref{eq:Split-Bregman_SH_a})
satisfies

\begin{equation}
\mathbf{R}^{\dagger}\left(\boldsymbol{y}-\mathbf{R}\boldsymbol{x}\right)+\frac{\mu}{\kappa}\left[\mathbf{S}^{\dagger}\left(\boldsymbol{q}^{\left(k\right)}-\mathbf{S}\boldsymbol{x}^{(k+1)}-\boldsymbol{b}^{\left(k\right)}\right)\right]=0,\label{eq:Zeros_Split-Bregman_a}
\end{equation}
where $\mathbf{R}^{\dagger}$ and $\mathbf{S}^{\dagger}$ are the
transposes of the matrices $\mathbf{R}$ and $\mathbf{S}$, respectively.
After regrouping we get ($\mathbf{I}$ is the $J\times J$ identity
matrix):

\begin{equation}
\left(\mathbf{R}^{\dagger}\mathbf{R}+\frac{\mu}{\kappa}\mathbf{I}\right)\boldsymbol{x}^{(k+1)}=\mathbf{R}^{\dagger}\boldsymbol{y}+\frac{\mu}{\kappa}\mathbf{S}^{\dagger}\left(\boldsymbol{q}^{\left(k\right)}-\boldsymbol{b}^{\left(k\right)}\right).\label{eq:equalization}
\end{equation}

The matrix $\left(\mathbf{R}^{\dagger}\mathbf{R}+\frac{\mu}{\kappa}\mathbf{I}\right)$
is positive definite (and, hence, invertible), but it is very large.
For reasons of computational cost, it makes sense in practice to approximate
$\boldsymbol{x}^{\left(k+1\right)}$ by applying a fixed limited number
$M$ of iterations of the method of conjugate gradients \cite{HESTENES:1952a}
to solve (\ref{eq:equalization}) for $\boldsymbol{x}^{(k+1)}$. A
single iterative step of the method is provided by the procedure \textbf{\textsc{Conjugate\_Gradients}}$\left(\mathbf{P},\boldsymbol{u},\boldsymbol{v},\boldsymbol{w},\boldsymbol{u'},\boldsymbol{v'},\boldsymbol{w'}\right)$,
where $\mathbf{P}$ is a $J\times J$ matrix, $\boldsymbol{u},\boldsymbol{v},\boldsymbol{w}$
are $J$-dimensional input vectors and $\boldsymbol{u'},\boldsymbol{v'},\boldsymbol{w'}$
are $J$-dimensional output vectors. Following \cite[p. 231]{HERMAN:2009a},
the details of this procedure are:

\begin{algorithmic}[1]

\STATE{\textbf{procedure }\textbf{\textsc{Conjugate\_Gradients}}$\left(\mathbf{P},\right.$$\boldsymbol{u},$$\boldsymbol{v},$$\boldsymbol{w},$$\boldsymbol{u'},$
$\boldsymbol{v'},$ $\left.\boldsymbol{w'}\right)$}

\STATE{~~~~\textbf{set} $\gamma\leftarrow\frac{\left\langle \boldsymbol{v},\boldsymbol{v}\right\rangle }{\left\langle \boldsymbol{w},\mathbf{P}\boldsymbol{w}\right\rangle }$}

\STATE{~~~~\textbf{set} $\boldsymbol{u'}\leftarrow\boldsymbol{u}+\gamma\boldsymbol{w}$}

\STATE{~~~~\textbf{set} $\boldsymbol{v'}\leftarrow\boldsymbol{v}-\gamma\mathbf{P}\boldsymbol{w}$}

\STATE{~~~~\textbf{set} $\delta\leftarrow\frac{\left\langle \boldsymbol{v'},\boldsymbol{v'}\right\rangle }{\left\langle \boldsymbol{v},\boldsymbol{v}\right\rangle }$}

\STATE{~~~~\textbf{set} $\boldsymbol{w'}\leftarrow\boldsymbol{v'}+\delta\boldsymbol{w}$}

\STATE{\textbf{end} \textbf{procedure}}

\end{algorithmic}

If we wish to approximate a solution of the system of equations $\mathbf{P}\boldsymbol{u=z}$,
then this can be achieved by first setting $\boldsymbol{u}^{(0)}$
to an arbitrary $J$-dimensional vector, $\boldsymbol{v}^{(0)}$ and
$\boldsymbol{w}^{(0)}$ to $\boldsymbol{z}-\mathbf{P}\boldsymbol{u}^{(0)}$
and then repeatedly iterating using \textbf{\textsc{Conjugate\_Gradients}}$\left(\mathbf{P},\boldsymbol{u}^{(k)},\boldsymbol{v}^{(k)},\boldsymbol{w}^{(k)},\boldsymbol{u}^{(k+1)},\boldsymbol{v}^{(k+1)},\boldsymbol{w}^{(k+1)}\right)$.
For a sufficiently large $K$, $\boldsymbol{u}^{(K)}$ will be a good
approximation to the desired solution.

In \cite{GOLDSTEIN:2009a} the solution to (\ref{eq:Split-Bregman_SH_b})
makes use of a procedure that implements the so-called soft shrinkage
operation that, when applied to an $I$-dimensional vector, produces
an $I$-dimensional vector with a smaller $\ell_{1}$-norm. This procedure
is \textbf{\textsc{Soft\_Shrink$\left(\tau,\boldsymbol{a},\boldsymbol{c}\right)$}},
where $\tau$ is a nonnegative real-valued input parameter; $\boldsymbol{a}=\left(a_{0},a_{1},\cdots,a_{I-1}\right)^{T}$
is an $I$-dimensional input vector and $\boldsymbol{c}$ is an $I$-dimensional
output vector. We use $\left|a\right|$ to denote the magnitude of
the number $a$. The details of this procedure are: 

\begin{algorithmic}[1]

\STATE{\textbf{procedure }\textbf{\textsc{Soft\_Shrink}}$\left(\tau,\boldsymbol{a},\boldsymbol{c}\right)$}

\STATE{~~~~\textbf{for} $i\leftarrow0$ to $I-1$ \textbf{do}}

\STATE{~~~~~~~~~\textbf{if} $\left|a_{i}\right|>\tau$ \textbf{then}}

\STATE{~~~~~~~~~~~~\textbf{set} $c_{i}\leftarrow\left(1-\frac{\tau}{\left|a_{i}\right|}\right)a_{i}$}

\STATE{~~~~~~~~~\textbf{else}}

\STATE{~~~~~~~~~~~~\textbf{set} $c_{i}\leftarrow0$}

\STATE{\textbf{end} \textbf{procedure}}

\end{algorithmic}

Under the conditions stated for the parameters of \textbf{\textsc{Soft\_Shrink$\left(\tau,\boldsymbol{a},\boldsymbol{c}\right)$}},
it will always be the case that $\left\Vert \boldsymbol{c}\right\Vert _{1}\leq\left\Vert \boldsymbol{a}\right\Vert _{1}$,
with the inequality strict unless all components of $\boldsymbol{a}$
are zero or $\tau$ is zero. The use of the softshrink operator is
feasible because there is no coupling between the elements of $\boldsymbol{c}$.

Based on this we can now state the algorithm proposed in \cite{VANDEGHINSTE:2013a}.
The user-specified input parameters to this algorithm are the $\kappa$,
$\mu$, $M$ and $K$ (the number of iterations).

\begin{algorithm}[H]
\caption{\label{alg:Belgian_Split-Bregman}\textbf{\textit{\emph{Modified Split
Bregman Algorithm}}}}
\end{algorithm}
\vspace{-5mm}

\begin{algorithmic}[1]

\STATE{\textbf{set} $k\leftarrow0$}

\STATE{\textbf{set} $\boldsymbol{x}^{\left(0\right)}\leftarrow\mathbf{0}$}

\STATE{\textbf{set} $\boldsymbol{q}^{\left(0\right)}\leftarrow\mathbf{0}$}

\STATE{\textbf{set} $\boldsymbol{b}^{\left(0\right)}\leftarrow\mathbf{0}$}

\STATE{\textbf{while} $k<K$}

\STATE{~~~~~\textbf{set} $m\leftarrow0$}

\STATE{~~~~~\textbf{set} $\boldsymbol{u}^{\left(0\right)}\leftarrow\boldsymbol{x}^{\left(k\right)}$}

\STATE{~~~~~\textbf{set} $\boldsymbol{v}^{\left(0\right)}$$\leftarrow$$\mathbf{R}^{\dagger}\boldsymbol{y}$$+$$\frac{\mu}{\kappa}\mathbf{S^{\dagger}}\left(\boldsymbol{q}^{\left(k\right)}-\boldsymbol{b}^{\left(k\right)}\right)$$-$$\left(\mathbf{R}^{\dagger}\mathbf{R}+\frac{\mu}{\kappa}\mathbf{I}\right)\boldsymbol{u}^{\left(0\right)}$}

\STATE{~~~~~\textbf{set} $\boldsymbol{w}^{\left(0\right)}\leftarrow\boldsymbol{v}^{\left(0\right)}$}

\STATE{~~~~~\textbf{while} $m<M$}

\STATE{~~~~~~~~~~\textbf{call }\textbf{\textsc{\small{}Conjugate\_Gradients}}\textbf{$\left(\mathbf{R}^{\dagger}\mathbf{R}+\frac{\mu}{\kappa}\mathbf{I},\right.$$\boldsymbol{u}^{(m)},$$\boldsymbol{v}^{(m)},$$\boldsymbol{w}^{(m)},$}

~~~~~~~~~~~~~~~~~~~~~~~~~~~~~~~~~~~~~~~~~~~~~~~~~~~~~~\textbf{$\boldsymbol{u}^{(m+1)},$$\boldsymbol{v}^{(m+1)},$$\left.\boldsymbol{w}^{(m+1)}\right)$}}

\STATE{~~~~~~~~~~\textbf{set} $m\leftarrow m+1$}

\STATE{~~~~~\textbf{set} $\boldsymbol{x}^{\left(k+1\right)}\leftarrow\boldsymbol{u}^{\left(M\right)}$
}

\STATE{~~~~~\textbf{set} $\boldsymbol{a}\leftarrow\mathbf{S}\boldsymbol{x}^{(k+1)}+\boldsymbol{b}^{\left(k\right)}$}

\STATE{~~~~~\textbf{call }\textbf{\textsc{Soft\_Shrink$\left(\frac{1}{\mu},\boldsymbol{a},\boldsymbol{q}^{(k+1)}\right)$}}}

\STATE{~~~~~\textbf{set} $\boldsymbol{b}^{(k+1)}\leftarrow\boldsymbol{b}^{\left(k\right)}+\left(\mathbf{S}\boldsymbol{x}^{(k+1)}-\boldsymbol{q}^{\left(k+1\right)}\right)$}

\STATE{~~~~~\textbf{set} $k\leftarrow k+1$}

\STATE{\textbf{return} $\boldsymbol{x}^{\left(K\right)}$}

\end{algorithmic}

\begin{figure}
\centering

\includegraphics[scale=0.39]{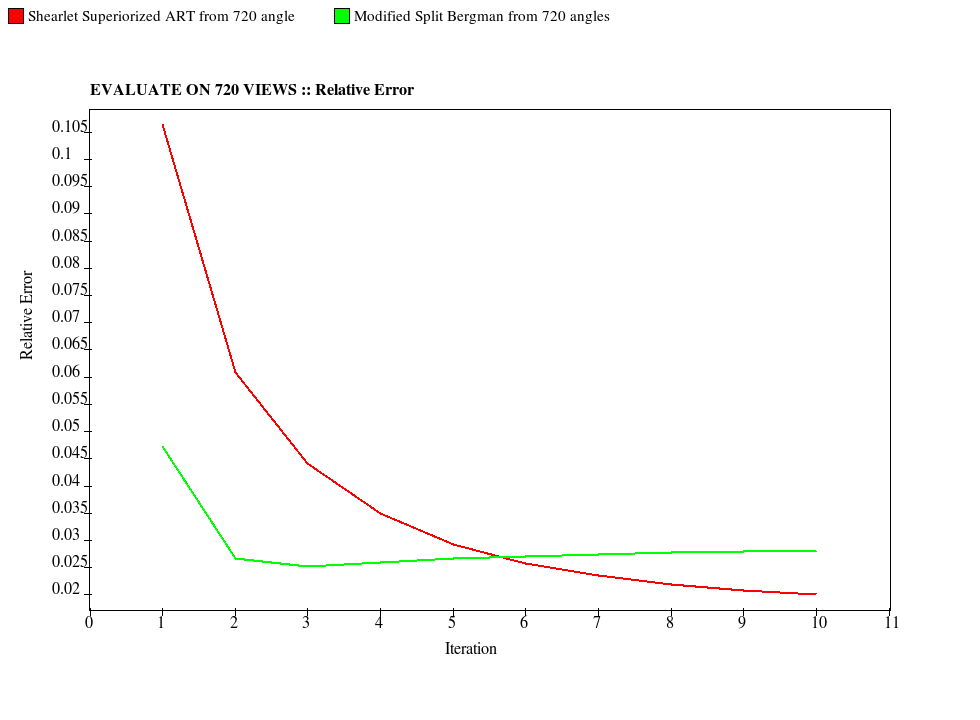}

\caption{\label{fig:Relative_Error_SBSH_SupSH}The relative errors (\ref{eq:relative_error})
of the reconstructions produced by the Shearlet-Based Superiorized
Version of ART and the Modified Split Bregman algorithms for the data
set with 720 projections.}
\end{figure}

\begin{figure}
\noindent \begin{centering}
\begin{tabular}{cc}
\subfloat[]{\label{fig:ARTS_IT_03}\includegraphics[scale=0.18]{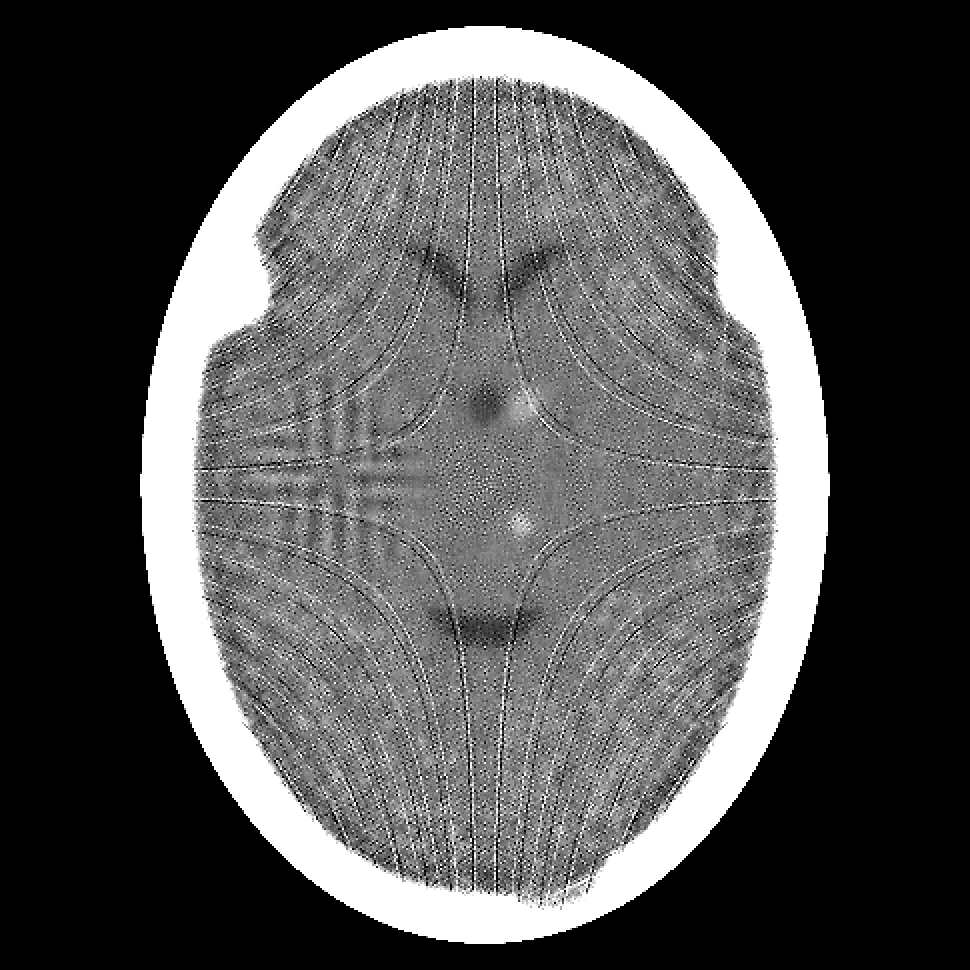}} & \subfloat[]{\label{fig:SBSH_IT_03}\includegraphics[scale=0.18]{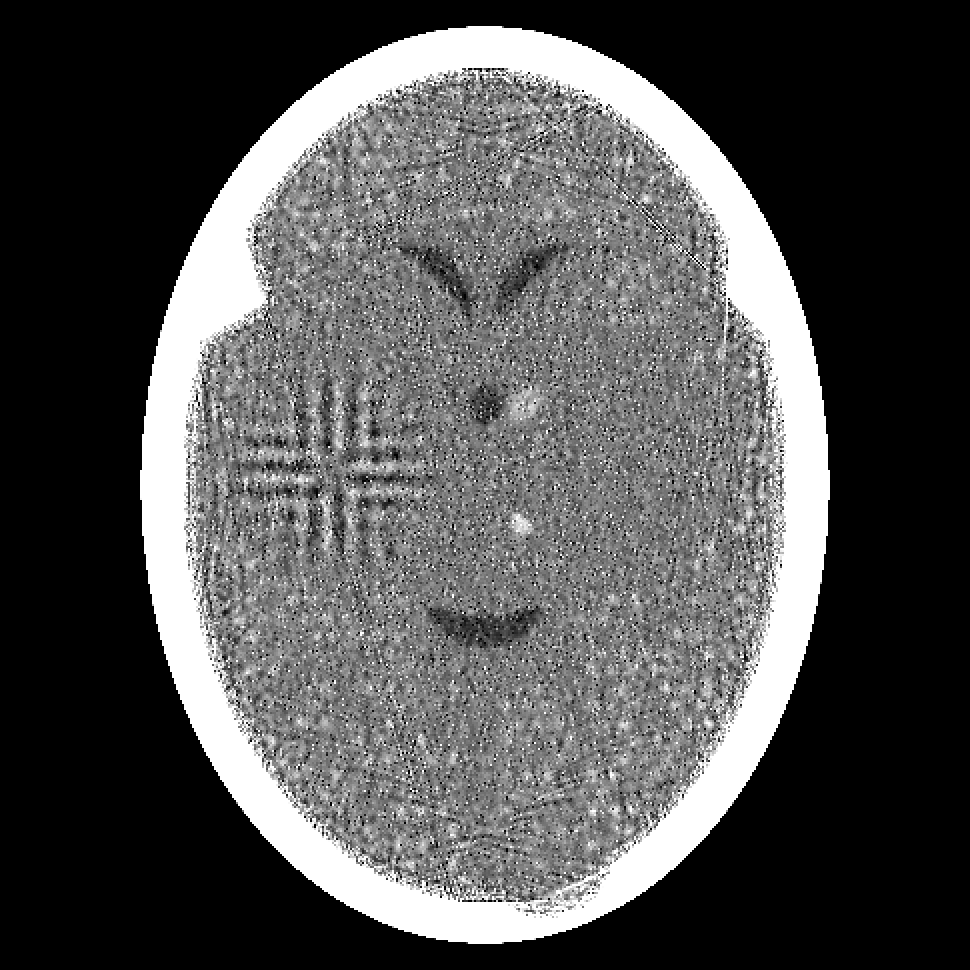}}\tabularnewline
\subfloat[]{\label{fig:ARTS_IT_10}\includegraphics[scale=0.18]{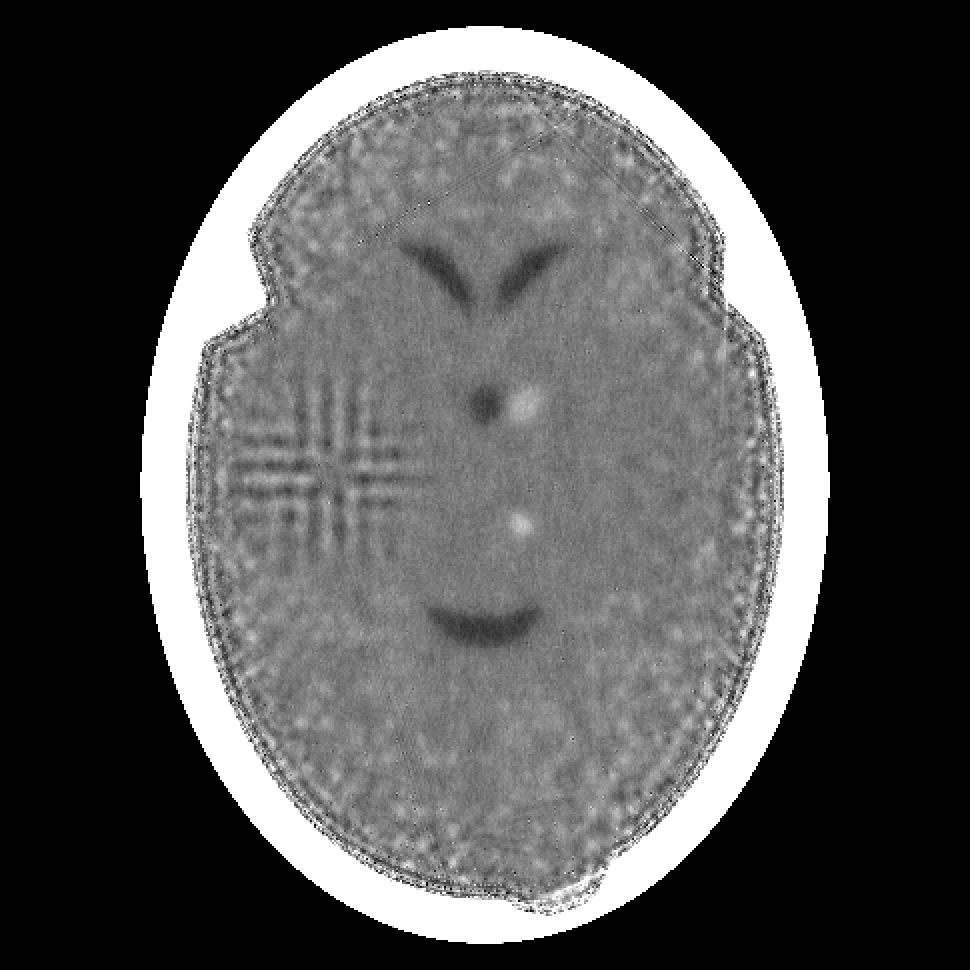}} & \subfloat[]{\label{fig:SBSH_IT_10}\includegraphics[scale=0.18]{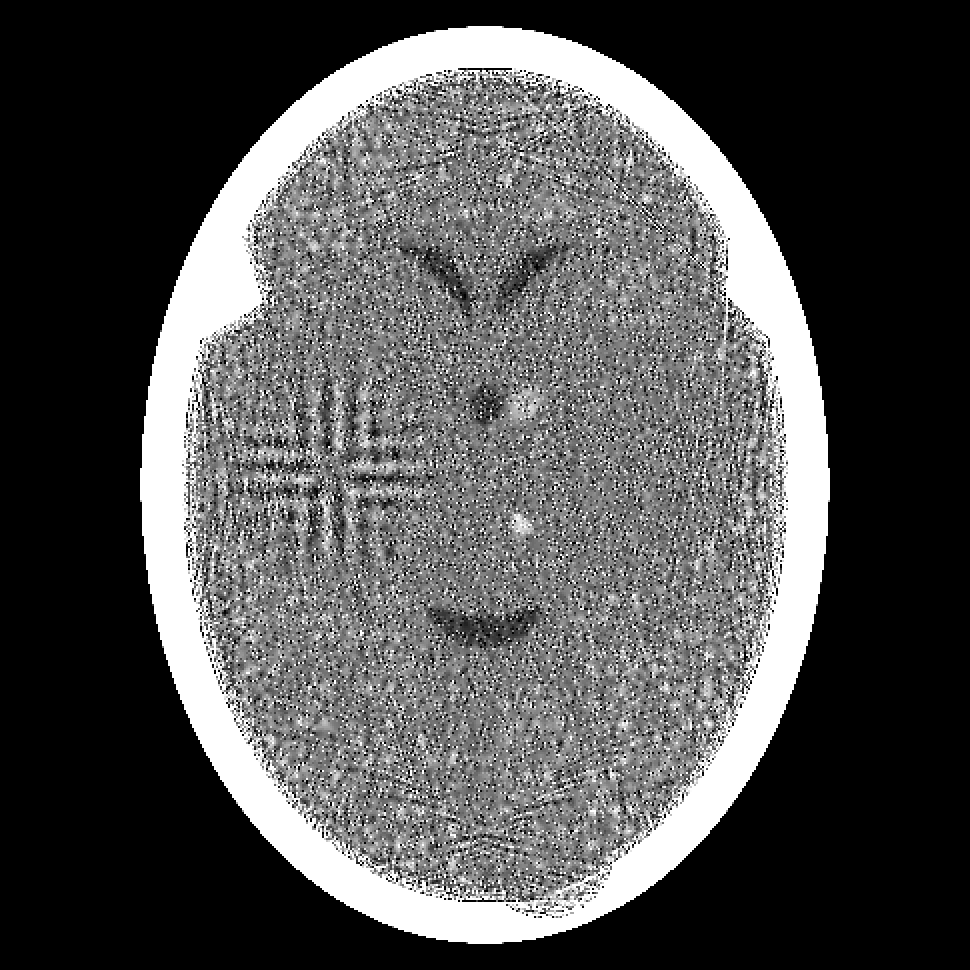}}\tabularnewline
\end{tabular}
\par\end{centering}
\centering{}\caption{\label{fig:Results_SBSH_ARTS}Results of the Shearlet-Based Superiorized
Version of ART (\emph{left column}) and the Modified Split Bregman
Algorithm (\emph{right column}) after the third (\emph{top row}) and
the tenth (\emph{bottom row}) iteration from 720 projections.}
\end{figure}

We compared the performance of the Shearlet-Based Superiorized Version
of ART with that of the Modified Split Bregman Algorithm on the projection
data used in Subsection \ref{subsec:Experiments_Single-Comparison};
the relative errors are reported in Fig. \ref{fig:Relative_Error_SBSH_SupSH}.
In Fig. \ref{fig:Results_SBSH_ARTS} we show the results of the two
approaches after the third and the tenth iteration.

\subsection{Using Total Variation}

We do not discuss the use of the split Bregman algorithm for TV in
the same detail as we have done for $\ell_{1}$-norm of the shearlet
transform, instead we just refer to an earlier work in which the topic
is discussed. In \cite{cccdh10} there is a comparison (see pp. 1082-3)
of the performance of superiorization with that of a version of the
split Bregman algorithm (suggested by T. Goldstein and S. Osher).
In the reported experiment the performances of the two approaches
are very similar according to both the primary and the secondary criterion,
but superiorization reached its output four times faster than the
split Bregman algorithm.

\section{Discussion and Conclusions\label{sec:Conclusions}}

The work on which we report above was originally motivated by the
positive results in \cite{VANDEGHINSTE:2013a} on iterative CT reconstruction
using shearlet-based regularization. We were particularly interested
whether or not shearlet-based regularization is more efficacious than
TV-based regularization (with which we had earlier positive experience)
on the kind CT reconstructions problems the we have been using for
the testing of reconstruction algorithms. For the purpose of this
investigation we brought the two regularization approaches within
a single framework using the methodology of superiorization. Making
use of the general concept of the superiorization of an iterative
algorithm, we have obtained Algorithm \ref{alg:Superiorized_ART}
that can be used for both TV-based and shearlet-based regularization.
In that algorithm there is a need in Step \ref{Line:NAV} for obtaining
a nonascending vector for the regularizing function (be that based
on TV or shearlets, or whatever). Our actually implemented algorithms
for TV-based and for shearlet-based regularization differ only in
the procedures \textbf{\textsc{NonascendingTV}} and \textbf{\textsc{NonascendingShearlet}}
that return a nonascending vector for TV and for shearlets, respectively.
(The design of the second of these procedures is an original contribution
of this paper, which is immediately applicable to finding nonascending
vectors for any regularization criterion that is expressed as the
$\ell_{1}$-norm of a transform specified by the application of a
matrix.) Having developed this common framework, we could perform
experiments comparing outcomes that depend only on the choice of the
regularization, without any other differences in implementation and
experimental design. In experiments reported above, TV-based superiorization
turned out to be more efficacious than shearlet-based superiorization.

While this observation is quite firm based on the reported experiments,
it is only fair to point out that it may be the case that, in spite
of our best efforts, we have not succeeded to select the parameters
of the Shearlet-Based Superiorized Version in an optimal manner. (An
example of a possible improvement is to change the number scales in
the fast Non-Iterative Shearlet Transform from the four, as specified
near the end of Section \ref{sec:CT_Problem}, to five; we have not
carried out a thorough investigation of the consequences of all such
possible changes.) An extension of the Shearlet-Based Superiorized
Version of ART, which includes both a TV term and a shearlet term
(claimed to be beneficial as compared to using only either one of
the two terms) was investigated in \cite{VANDEGHINSTE:2012b}. In
our paper we compared the performance of algorithms in which just
one of the terms is used.

We also reported comparisons of our superiorization approaches with
two commonly-used CT reconstruction methods: FBP and (unsuperiorized)
ART, as well as with the use of the split Bregman approach. The outcome
of these comparisons is somewhat ambiguous. For example, TV-Based
Superiorized ART was found better than FBP by all reported measures,
except in the case when the number of projections is not very small
(i.e., 360 or more) and the medical task that provides the figure
of merit is the localization of very small tumors in the brain. On
the other hand, TV-Based Superiorized ART was found better than unsuperiorized
ART by all reported measures.

\section*{Acknowledgments{\normalsize{}\label{sec:Acknowledgments}}}

This work was supported in part by the National Science Foundation
Award No. DMS-1114901. We wish to thank Bert Vandeghinste and, particularly,
Bart Goossens (who has kindly provided a detailed critical review
of an earlier version of this paper) for their assistance and for
providing us with source code implementing the Discrete Shearlet Transform.
We also thank Marcelo Zibetti and Chuan Lin for comments on an earlier
version of this paper.

\section*{References{\normalsize{}\label{sec:References}}}

\end{document}